\documentclass[lettersize,journal]{IEEEtran}
\usepackage{amsmath,amsfonts}
\usepackage{algorithmic}
\usepackage{algorithm}
\usepackage{array}
\usepackage[caption=false,font=normalsize,labelfont=sf,textfont=sf,subrefformat=parens,labelformat=parens]{subfig}
\usepackage{textcomp}
\usepackage{stfloats}
\usepackage{url}
\usepackage{verbatim}
\usepackage{graphicx}
\usepackage{cite}
\usepackage{multirow}
\usepackage{booktabs}
\usepackage{adjustbox} 
\usepackage{tabularray}
\usepackage{nameref}
\usepackage{hyperref}
\hypersetup{colorlinks,linkcolor={blue},citecolor={blue},urlcolor={black}} 
\usepackage{orcidlink}

\begin{document}

\title{An Empirical Fault Vulnerability Exploration of ReRAM-based Process-in-Memory CNN Accelerators }

\author{Aniseh Dorostkar$^{\,\orcidlink{0000-0003-2596-0721}}$, Hamed Farbeh$^{\,\orcidlink{0000-0002-4204-9131}}$, Hamid R. Zarandi$^{\,\orcidlink{0000-0003-1385-4171}}$,~\IEEEmembership{member,~IEEE,}
\thanks{Manuscript received 07 March 2023; revised 11 March 2024; accepted 15 May 2024. (Corresponding
	author: Hamed Farbeh (Email: farbeh@aut.ac.ir).)}
\thanks{Aniseh Dorostkar, Hamed Farbeh, and Hamid R. Zarandi are with Amirkabir University of Technology (Tehran Polytechnic), Tehran, Iran (Email: a\_dorostkar@aut.ac.ir, farbeh@aut.ac.ir, h\_zarandi@aut.ac.ir) }
\thanks{Digital Object Identifier 10.1109/TR.2024.3405825}}

\markboth{IEEE Transactions on Reliability,~Vol.~74, No.~1, June~2024}%
{Shell \MakeLowercase{\textit{et al.}}: A Sample Article Using IEEEtran.cls for IEEE Journals}

\IEEEpubid{\begin{minipage}{\textwidth}\ \\[8pt]  \centering
		1558-1721~\copyright~2024 IEEE. Personal use is permitted, but republication/redistribution requires IEEE permission.\\ 
		See \url{https://www.ieee.org/publications/rights/index.html} for more information.
\end{minipage}}

\maketitle
\IEEEpubidadjcol
\begin{abstract}
Resistive random-access memory (ReRAM)-based \textit{Processing-in-Memory} (PIM) accelerator is a promising platform for processing massively memory intensive matrix-vector multiplications of neural networks in parallel domain, due to its capability of analog computation, ultra-high density, near-zero leakage current, and non-volatility. Despite many advantages, ReRAM-based accelerators are highly error-prone due to limitations of technology fabrication that lead to process variations and defects. These limitations degrade the accuracy of Deep Convolutional Neural Networks (Deep CNNs) running on PIM accelerators. While these CNNs accelerators are widely deployed in safety-critical systems, their vulnerability to fault is not well explored. In this paper, we have developed a fault injection framework to investigate the vulnerability of large-scale CNNs at both software- and hardware-level of inference phases. Faulty ReRAM devices are another reliability challenges due to significant degradation of classification accuracy when CNN parameters are mapped to the accelerators. To investigate this challenge, we map the CNN learning parameter to the ReRAM crossbar and inject faults into crossbar arrays. The proposed framework analyzes the impact of \textit{stuck-at high} (SaH) and \textit{stuck-at low} (SaL) fault models on different layers and locations of CNN learning parameters. By performing extensive fault injections, we illustrate that the vulnerability behavior of ReRAM-based PIM accelerator for CNNs is greatly impressible to the types and depth of layers, the location of the learning parameter in every layer, and the value and types of faults. Our observations show that different models have different vulnerabilities to faults. Specifically, we show that SaL further reduces classification accuracy than SaH. 
\end{abstract}


\section{Introduction}
\IEEEPARstart{I}{n recent} years, deep neural networks (DNNs) and particularly convolutional neural networks (CNNs) have been increasingly deployed in many fields, including safety-critical systems, e.g., healthcare, autonomous vehicles, and  space applications~\cite{ibrahim2020soft}. The need for high classification accuracy leads to an increase in the complexity of CNN models and the number of parameters as well as the  computation demand of neural networks. Meanwhile, energy consumption and memory/storage occupation significantly increased. As CNN depth increases, computational/memory resources and interconnections face severe challenges, e.g., there are about 635,200 interconnections in a fully-connected two-layer deep CNN to classify the MNIST dataset~\cite{lecun1998gradient}. VGG~\cite{simonyan2014very} as another well-known CNN model has the order of million parameters~\cite{xu2021advances}. Conventional CPUs and Graph Processing Units (GPUs) are not able to efficiently process such a large-scale computation with enormous memory resources owing to power wall and memory wall bottlenecks. A novel way to tackle these challenges is the  processing-in-memory (PIM) approach, which moves a majority of computation into memory and augments the memory to execute computing tasks directly within its arrays. Conventional memory technologies such as dynamic random-access memory (DRAM) and emerging non-volatile memories such as spin-transfer torque magnetic RAM (STT-MRAM), resistive random-access memory (ReRAM), and phase change memory (PCM), are capable of doing logical and arithmetic computations beyond data storage~\cite{chi2016prime}. Among these technologies, ReRAM has advantages of higher density, lower power consumption, higher speed, and better scalability. Moreover, the capability of operating in both analog and digital domains and high-density storage makes this technology a promising candidate for accelerating big data processing and increasing the efficiency of CNNs computations.

The analog-based PIMs~\cite{shafiee2016isaac, 10051721_XCRYPT} perform parallel Vector-Matrix Multiplications (VMM) using the natural multiply-and accumulate feature of ReRAM crossbars, which are extremely energy efficient, have a lower delay, and are less accurate than non-analog PIMs. Meanwhile, non-analog PIMs~\cite{9407200_Prasad} perform massively parallel Boolean operations, which compromises accuracy for energy efficiency and delay.  
\IEEEpubidadjcol

ReRAM-based PIM accelerator is a promising platform for processing neural networks. Due to the characteristics of neural networks, existing ReRAM-based accelerators, e.g., ISAAC~\cite{shafiee2016isaac}, PRIME~\cite{chi2016prime}, ZARA~\cite{chen2019zara}, PipeLayer~\cite{song2017pipelayer}, and FORMS~\cite{FORMS_9499856_Yuan}, can reduce time and energy consumption compared to their CMOS-based counterparts and von-Neumann processors. These improvements are due to the crossbar architecture of ReRAM, which can perform vector-matrix operations in the analog domain and eliminate transmission of neural network data such as weights between the computing unit and memory~\cite{ghasemi2022grapha}. The crossbar array of ReRAM can perform neural network’s main operations, i.e., VMM and reduce the time complexity of VMM operations from O($n^2$) to O($n$) and even in the best case to O($1$)~\cite{lammie2020memtorch}.

Despite the mentioned advantages, ReRAM-based accelerators are highly error-prone due to the limitations of technology fabrication. Accordingly, they suffer from Process Variations (PVs), limited write endurance, the discrete and limited number of conductance states, transient and permanent faults, sources of noises, and non-linearity of current-voltage characteristics~\cite{lammie2020memtorch, li2019improving, chakraborty2020geniex}. ReRAM cells are susceptible to many types of hardware faults both in the manufacturing and operational phases. Considering the capability of employing ReRAM as both storage and logic elements, its storage application is more sensitive to faults than their logical counterpart. Variations in the resistance of cells directly cause error occurrence in the result of VMM and reduction of computational accuracy. Therefore, PVs and defects in fabrication process of ReRAM technology not only reduce the yield of the memory devices, but also degrade the accuracy of CNNs running on the PIM-accelerators~\cite{song2020itt}. Learning parameters of neural networks, represented as the conductance of ReRAM memory cells, can be easily altered by faults in ReRAM technology. 

CNN inference application continuously processes many safety-critical inference tasks where their correctness is more important than its training phase in this domain. Even one single fault in the ReRAM crossbar causes error occurrence in the inference application that can affect the reliability of whole system. Therefore, to design a CNN model for safety-critical systems, this makes it necessary to explore the behavior of each layer against fault to guarantee designing a robust model. It is necessary to investigate the vulnerability of CNN models based on ReRAM-based PIM accelerator characteristics. 

A few prior studies analyzed CNNs reliability and vulnerability to permanent and transient faults based on several CNNs for different hardware platforms, e.g., CPUs, GPUs, Tensor Processing Units (TPUs), Field Programmable Gate Arrays (FPGAs), and Application-Specific Integrated Circuits (ASICs)~\cite{ibrahim2020soft, gambardella2019efficient, zhang2019fault }. Nevertheless, still, no attention was paid to investigating the vulnerability of ReRAM-based PIM accelerators for CNN-type algorithms. In addition, there are some studies on the sensitivity of CNNs to changing their learning parameters and occurrence of the faults in neurons, weights, or data path of CNNs~\cite{li2017understanding, mahmoud2020pytorchfi, reagen2018ares, neggaz2019cnns, hari2017sassifi, tsai2021nvbitfi, chen2020tensorfi}. None of the previous works investigated the effects of changing the value of learning parameters in a specified location and perusing permanent faults, such as stuck-at maximum and minimum values separately in weights and kernels of CNNs. 

In this paper, the vulnerability of different CNN models to faults and the causes of faults in ReRAM crossbar arrays as well as their effects on the classification accuracy of CNNs, are investigated. We have demonstrated that decreasing the classification accuracy of CNNs highly depends upon the location of corrupted weights or, on the other hand, in which crossbars the corrupted weights are mapped. We have developed a framework for fault injection into both software- and hardware-level of CNNs accelerators. At the software-level, the vulnerability of the CNN models is explored by changing the hardware-independent learning parameters or without considering the source of faults. At hardware-level fault injection, the vulnerability of CNN models is investigated by considering the hardware specification, and permanent faults are injected into ReRAM cells of the accelerator. Hence, we deeply investigate the wide range of diversity changes in the learning parameters of CNN models with different topologies. We also examine the vulnerability of different layers in several models of CNNs with specific topologies. The main contributions of this paper are as follows:

\begin{enumerate}
\item{Both single and multiple faults are injected into the software-level of CNN models to investigate and study their vulnerability to different fault behaviors. We comprehensively analyze the behavioral simulation of stuck-at faults by injecting maximum and minimum values to investigate the vulnerability of CNNs against to stuck-at high resistance (SaH) and stuck-at low resistance (SaL) faults respectively.}
\item{Hardware-level fault injection for ReRAM-based PIM accelerator is also performed. It helps to investigate the effect of hardware faults on the output of CNNs. To this aim, we map the learning parameters of the inference phase as conductance of ReRAM crossbars and then inject SaH and SaL faults into crossbar cells.}
\item{Through extensive simulations and analysis, we reveal that the output of large-scale CNNs highly depends on the type and value of faults, topology models, types and the number of layers, and distance of layers from output.}
\item{By performing extensive fault injection experiments, we illustrate that the behavior of SaH and SaL faults differently affects the output of CNNs, and the degradation in classification accuracy due to the occurrence of SaL is significantly higher than that of SaH.}
\end{enumerate}

The rest of this article are organized as follows. Section~\ref{Preliminaries} provides background on neural networks, ReRAM devices, and reliability factors of ReRAM-based CNN accelerators. The complete analyses and results of the vulnerability investigation are provided in Section~\ref{The Proposed Framework for Reliability Investigation in CNNs}. In Section~\ref{Summery and Discussion}, the summary of investigations and simulation results and detailed analysis are provided. A comprehensive discussion of related work is provided in Section~\ref{Related Work}. Finally, Section~\ref{Conclusion} concludes this article.  

\section{Preliminaries}\label{Preliminaries}
\subsection{Neural Networks}
As one of the most prominent Artificial Neural Networks (ANNs), Convolutional Neural Networks (CNNs) have been made of convolutional layers (\textit{CONV}), pooling layers (\textit{POOL}), fully-connected layers or classification layers (\textit{FC} or \textit{DENSE}), and some other custom layers like BatchNorm [6]. These layers, have some parameters, e.g., weights, kernel or filter, and bias, that must be adjusted and learned in the training phase and are called \textit{learning parameters}. CNNs usually consist of a group of \textit{CONV} and \textit{POOL} layers sequentially placed at that end with some \textit{FC} or classification layers (usually less than three layers). The main computations performed in CNNs contain multi-dimensional convolutional and VMM. In \textit{FC} layers, VMM operation is performed in each neuron, by the following equation:
\begin{equation}
\label{eq1}
y_{j} = \sum_{i=0}^{n}x_{i}  w_{ij} + b_{j}
\end{equation}
where $y_j$ is the output of $j^{th}$ neuron, $x_i$ is the input of $j^{th}$ neuron (output of $i^{th}$ neuron in the previous layer), $w_{ij}$ is the weighted synapses corresponding to each input, and $b_{j}$ is the bias of each neuron. The result is given to the activation function of $\sigma$, which can be either \textit{sigmoid}, \textit{ReLU}, or \textit{tanh}, and the output of the neuron is calculated. In addition, a convolutional operation in \textit{CONV} layers is performed for each kernel and its corresponding Input Feature Map (IFM) to extract the Output Feature Map (OFM) as
\begin{multline}
	\label{eq2}
	y_{j} =\sigma\sum_{i=0}^{C_{in}}\big(\sum_{dx=0}^{C_{W}}\bigg(\sum_{dy=0}^{C_{H}}	IFM(i,x+dx,y+dy)\big){*}\\K_{n}(i,dx,dy)\bigg)+B_{i},1\leq n \leq C_{out}
\end{multline}
where $IFM$ is the input feature map, $C_{in}$ is the number of feature maps, $C_{out}$ is the number of output feature maps, $K$ is the kernel matrix of \textit{CONV} layers, $K_{n}(i,dx,dy)$ is the $n^{th}$ filter matrix for input $i^{th}$ layer, and $B_i$ is the bias array of each kernel.

\subsection{ReRAM Memory and Crossbar Structure}
\begin{figure}[!t]
	\centering
	\subfloat[\label{fig1-a}]{
		\includegraphics[width=0.65\linewidth]{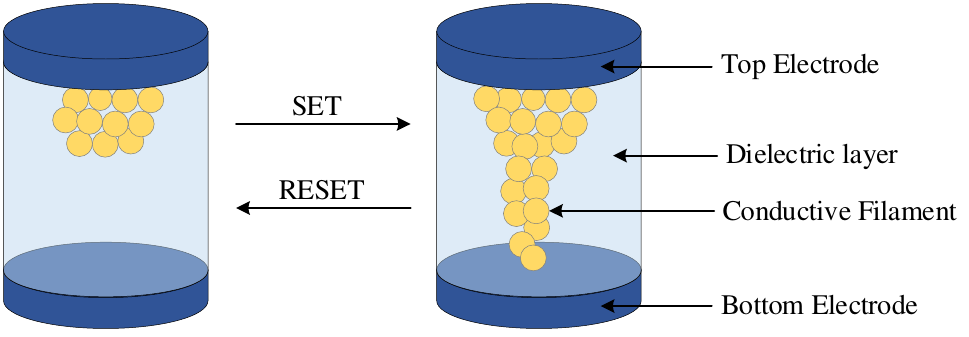}
	}
	\subfloat[\label{fig1-b}]{
		\includegraphics[width=0.35\linewidth]{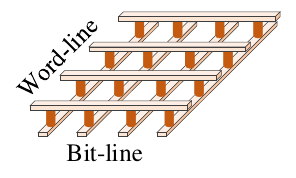}
	}
	\caption{ReRAM structure. a) ReRAM cell and b) ReRAM crossbar structure.}
	\label{fig1:multi2}
\end{figure}

ReRAM is a nonvolatile memory that stores data with variable resistant states. ReRAM is constructed with three layers containing a metal-insulator oxide (dielectric) layer sandwiched between an electrochemically active metal top electrode and an electrochemically inert metal bottom electrode. The schematic of ReRAM is shown in Fig.~\ref{fig1-a}. By controlling the external ‘set’ and ‘reset’ voltage, the state of the cell is switched between the low resistant state (LRS or logical ‘1’) and the high resistant state (HRS or logical ‘0’), which is caused by the formation and rupture of the conductive filament in the dielectric layer. To read data from the ReRAM cell, a small voltage referred to as the read voltage is applied to identify whether the state of the cell is in the HRS (logical ‘0’) or the LRS (logical ‘1’) state. This low voltage will not disturb the current state of the cell. 

Using multiple cells of ReRAM, a ReRAM crossbar structure is constructed, as shown in Fig.~\ref{fig1-b}. By storing matrix elements as conductance on the ReRAM crossbar array, and applying a vector as input voltage signals, the ReRAM crossbar can perform matrix-vector multiplication in the analog domain.

1R (one-resistor) and 1T1R (one-transistor-one-resistor) crossbar structures are two commonly used structures for ReRAM crossbars. In 1R structure, each ReRAM cell consists of one resistor as a storage element whereas in 1R1T structure, each cell consists of an access transistor to select and control each cell and a resistor as a storage element. However, 1R structure is simpler than other structures and has a higher density. 1T1R structure has attracted significant attention due to its ability to reduce the sneak-path current issue happening in ReRAM crossbars. Moreover, in recent decades, 1T1R structures have also been used for implementing emerging applications such as PIM architectures and neuromorphic computing~\cite{shafiee2016isaac, chi2016prime, sun2021unary, Wang_2022, 7446049_Bojnordi, FORMS_9499856_Yuan}.
  
\begin{figure}[!t]
	\centering
	\includegraphics[width=0.5\textwidth]{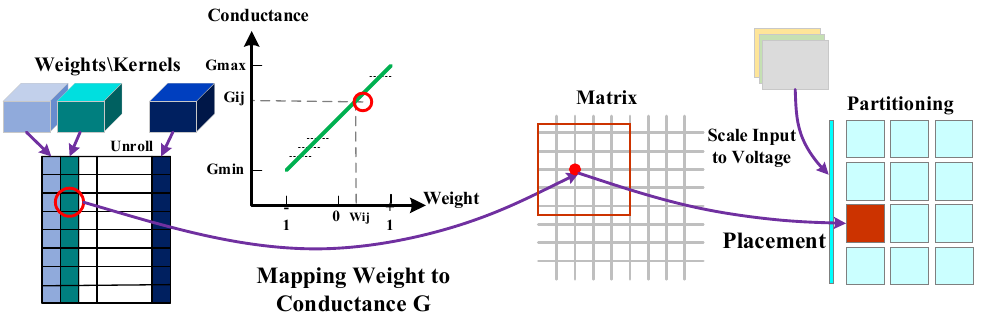}
	\caption{General process of mapping and placement of learning parameters to ReRAM crossbar.}
	\label{fig2-mapping}
\end{figure}

The general process of mapping and placement of learning parameters, e.g., weight or kernel matrix and bias vector, on ReRAM-based PIM accelerators are shown in Fig.~\ref{fig2-mapping}. At first, the learning parameters, including the weight matrix and bias vector of FC layers as well as kernel and bias vector of convolutional layers, must be converted to an appropriate matrix to perform VMM in crossbar architecture correctly. Then the converted matrix is mapped to the corresponding conductance value. Since the number of parameters and size of the matrix varies in each layer, the size of the converted matrix may be smaller/larger than the crossbar size. Therefore, based on the CNN model, an adequate placement mechanism should be done in a hierarchy of crossbars, tiles, and memory banks of the ReRAM accelerator.

\begin{figure}[!t]
\centering
\includegraphics[width=0.35\textwidth]{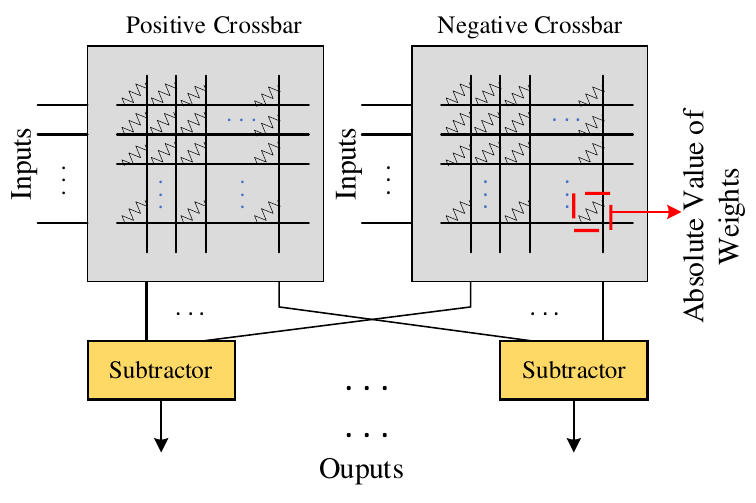}
\caption{Dual crossbar architecture to store positive and negative values.}
\label{fig3-dual-crossbar}
\end{figure}

In general, learning parameters contain both positive and negative values whereas, the ReRAM cell stores the absolute values of the parameters as conductance. To solve this challenge, and map both positive and negative values, two crossbars containing a positive crossbar and a negative crossbar have been proposed~\cite{chi2016prime, sun2021unary} (shown in Fig.~\ref{fig3-dual-crossbar}). The absolute value of learning parameters is stored in the negative/positive crossbar and their corresponding cells in dual crossbar maps with $G_{min}$. The conductance value of each cell is calculated by the following
equation:
\begin{align}\label{eq3}
{G^{*}_{i,j,\text{pos}}} &= \begin{cases}
	G_{i,j},&{\text{if }} w_{i,j}\geq{0} \\ 
	{G_{min},}&{\text{if }} w_{i,j}<{0}
\end{cases},\nonumber \\
{G^{*}_{i,j,\text{neg}}} &= \begin{cases}
	|G_{i,j}|,&{\text{if }} w_{i,j}\leq{0} \\ 
	{G_{min},}&{\text{if }} w _{i,j}>{0}
\end{cases}
\end{align}

where $G_{(i,j)}$ is the base conductance that must be mapped to ReRAM cells and $G_{i,j,pos}^*$ and $G_{i,j,neg}^*$ are the absolute conductance values that map to positive and negative crossbars, respectively. 

\subsection{Reliability of ReRAM-based Accelerators for CNNs}
There is no comprehensive evaluation criterion for the reliability of the ReRAM-based analog PIM accelerator for processing CNNs. The classification accuracy of CNNs can be considered as a parameter to quantify the reliability criterion of a neural network~\cite{zhao2020reliability}. In a ReRAM crossbar array, when the matrices are mapped into the conductance of the crossbar array and input values are scaled to the voltage of inputs, the output current value of each column is not exactly the same as the sum of all currents passing through each cell, however it is slightly different from its ideal value. In reality, due to non-ideal sources, this current value differs from the expected current value which itself causes an error in analog computations~\cite{yeo2019stuck}, leading to an error in the weight connected to neurons. Non-ideal behaviors have a significant effect on VMM calculation, and this itself is intensified due to the faults of the ReRAM cells~\cite{chakraborty2020geniex, li2019improving}. Consequently, non-ideal behaviors significantly reduce the classification accuracy of the neural network mapped onto the ReRAM accelerator. 

Various factors can affect the classification accuracy and reliability of ReRAM-based accelerators. In general, these factors can be divided into two categories: 1) data-dependent factors and 2) data-independent factors. Data-dependent factors are related to the values of memory cells, their variations, and the way of conversion and calculation. These factors include the way of mapping weights or kernels and bias into conductance in the ReRAM crossbar and scaling the input data into voltage, the non-ideal behavior of ReRAM cells caused by PVs (cell-to-cell variations and cycle-to-cycle variations), resistance drift, quantization error, and the presence of faults (such as stuck-at faults). Data-independent factors are mainly caused by peripheral circuits and fabrication processes, e.g., access transistor or selector, Analog-To-Digital Converters (ADC), Digital-To-Analog Converters (DAC), resolution and sneak path current, and parasitic resistance~\cite{charan2020accurate}. 

High fault rates and PVs lead to accelerator reliability degradation. Meanwhile, to implement an extensive computing system with ReRAM-based PIM, the wire density becomes significantly high, and the reliability of read and write operations decreases due to voltage drop and PVs. As the rate of ReRAM faults increases, the accuracy of the neural network decreases drastically.

Learning parameters of neural networks, represented as the conductance of ReRAM memory cells, can be easily changed by the limitations of ReRAM technology. For example, PVs as well as thermal and electrical noise can limit the programming accuracy of ReRAM cells. The cell resistance variations directly affect the VMM results, causing errors and reducing the classification of CNN’s accuracy~\cite{xia2017mnsim}. Resistance drift disturbs the state of the memory cells and, in turn, it reduces the accuracy of calculations, efficiency, and stability of the system. Therefore, it is necessary to analyze and investigate the effects of faults in CNNs  carefully. To this aim, the effect of fault sources in CNN models as well as ReRAM-based PIM accelerators should be investigated.

The manufacturing process of ReRAM is not so mature as that of CMOS. As a result, ReRAM cells are susceptible to many types of hardware faults in both the manufacturing and  data processing phases. ReRAM faults are divided into permanent and transient faults based on their temporal characteristics~\cite{gambardella2019efficient}. Transient faults may exist for a short period and are usually caused by external disturbances such as high energy particles hitting the electrical device (such as randomly changing cell values), temperature, resistance drift, process variations, and write/read disturbance~\cite{ibrahim2020soft, xia2017fault, cheshmikhani2016investigating, cheshmikhani2019system}. Transient faults in ReRAM memory change the resistance of the cells without physically damaging the cells. Read disturbances in ReRAM memory are considered as transient faults~\cite{xia2017stuck, chen2014rram}. Permanent faults are continuous and remain unchanged over time. These faults are usually due to irreversible physical damage such as limitation of manufacturing methods at the fabrication stage~\cite{chen2014rram}, e.g., oxide layer thickness, fluctuation, and decreasing feature size~\cite{yeo2019stuck}, or limitation in endurance~\cite{beckmann2016nanoscale, grossi2019resistive}. When the cell is permanently faulty, the resistance of the ReRAM cell will not change, thus causing the write operation to fail. Defects in manufacturing time such as short defects and over-forming stage defects~\cite{chen2014rram} cause the resistance of the cell to become stuck-at high (SaH) or stuck-at low (SaL). The metal oxide layer of ReRAM is sensitive to fabrication thickness and forming method, and it is challenging to avoid stuck-at faults during the fabrication stage [16]. Even though the ReRAM crossbar array passes the manufacturing test, faults can also happen during the time in use and data processing\cite{grossi2019resistive}. These faults can be attributed to the limitation of ReRAM writes endurance or to the use of several consecutive write-0 (or write-1) pulses~\cite{mittal2018survey}. SaH usually occurs due to short defects that cause part of the cell always to have high resistance. SaL occurs due to permanent cell failure or a broken selector or wire. Stuck-at faults are randomly distributed across the crossbar array~\cite{yeo2019stuck}.

In neural networks, the fault model is defined just based on the behavior of the neuron and independent of its physical implementation. Finally, a digital platform is used to investigate the effect of a fault on neural computations. In the binary implementation, these two faults are modeled as stuck-at faults (SaL and SaH). Synapse faults are modeled as stuck-at values (SaV), where V is a number between [$w_{min}$, $w_{max}$] where $w_{min}$ and $w_{max}$ are minimum and maximum values of weights, kernels and biases, respectively. Neuron body faults are typically saturated to positive or negative values modeled by Sa1 or Sa(-1) since the output value of the activation function is in this range. Stuck-at faults allow modeling faults at the behavioral level, independent of the implementation platform and the precise characteristics of the physical faults~\cite{mittal2018survey}. 
 
Since the processing unit of ReRAM accelerators is based on a crossbar array that performs multiplication and addition operations concurrently, the possibility of faults occurring in ReRAM cells is more significant than in conventional ReRAM memory/storage. Hence, the main focus of this research is on the presence of faults in ReRAM cells. For this purpose, in the following, faults in the synaptic weights of the CNNs and the conductance of ReRAM cells will be investigated and analyzed. The faults are modeled based on SaV, SaH, and SaL.
  \begin{figure*}[!t]
 	\centering
 	\includegraphics[width=0.7\textwidth]{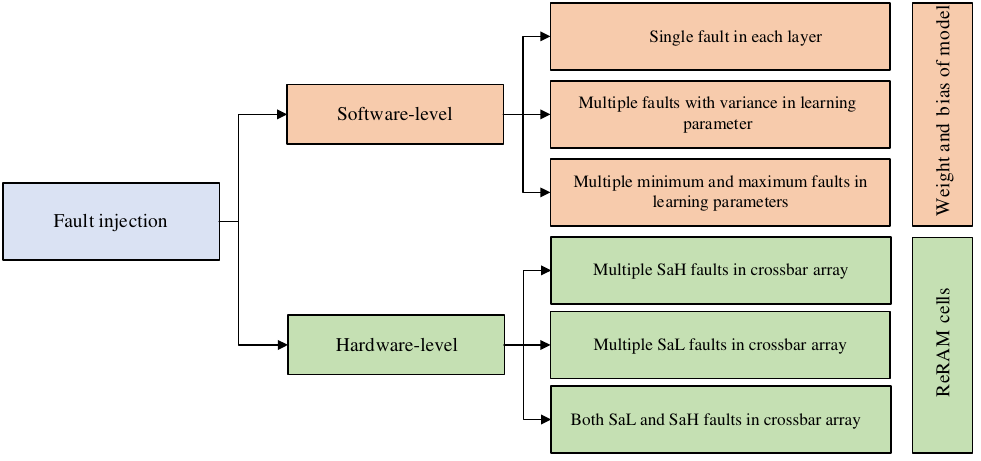}
 	\caption{Different phases of the proposed ReConFI faults injection framework.}
 	\label{fig4-different phases}
 \end{figure*}
\section{ReConFI: The Proposed Framework for Reliability Investigation in CNNs}\label{The Proposed Framework for Reliability Investigation in CNNs}
The proposed fault injection framework, called \textit{\underline{Re}RAM \underline{Con}volutional Neural Network \underline{F}ault \underline{I}njector} (ReConFI), performs fault injection at both software- and hardware-level and monitors the effect of faults. At software-level fault injection, the CNNs learning parameters are corrupted to study the CNN models behavior in the presence of faults without considering the hardware platform specification. In hardware-level fault injection, the hardware specification of the accelerator is considered, and the faults are injected into ReRAM cells and the conductance of cells which denotes as learning parameters of CNN models are changed. Then, the effects of changing conductance on the output of CNN models are obtained.

To investigate the behavior of faults in CNNs and analyze their vulnerability, a three-phase fault injection has been performed at the software-level. In the first phase, single faults are injected into the learning parameters of weights, kernels, and biases at the algorithm-level without considering the processing platform. Ten million fault injection experiments, consisting of 1000 faults for 10000 input images, are injected at random locations of weights, kernels, and biases in each layer. In the second phase of fault simulation in CNN models, the focus is on stuck-at value (SaV) faults, in such a way that different numbers of weights, kernels, and biases are stuck-at a specific value. Multiple faults with the magnitude of 0.05x to 10x of golden value are added to the corresponding learning parameters. In the third phase, to investigate the vulnerability of each layer against stuck-at faults, SaH and SaL faults are modeled in synapsis weights at the software-level. More than one fault with the value of the maximum and minimum amount has been injected into the weights, kernels, and bias matrices. The maximum and minimum values indicate SaH resistance and SaL resistance, respectively.

In the hardware-level fault injection, learning parameters are mapped to the conductance of cells, and inputs are scaled to the voltage range. Then, SaH and SaL faults are injected into the ReRAM-based accelerator platform after mapping the learning parameters of CNN to the conductance of ReRAM cells. Multiple SaH faults, and SaL faults, and their combinations are separately injected into ReRAM cells that contain the value of learning parameters. Different phases of the proposed fault injection framework are shown in Fig.~\ref{fig4-different phases}.

\subsection{Simulation Platform and Neural Network Models}
Different neural networks have different topologies, the type of which can affect error propagation. In this paper, fault injection has been done for different types of CNNs, and fault injection is investigated for different layers. Moreover, different CNNs have different layers, in terms of position, type, number of each type, and the total number of layers. The occurrence of a fault in the weight or kernel of each layer can have a different effect on the output and accuracy of the network. 

\begin{figure*}[!t]
	\centering
	\includegraphics[width=0.85\linewidth]{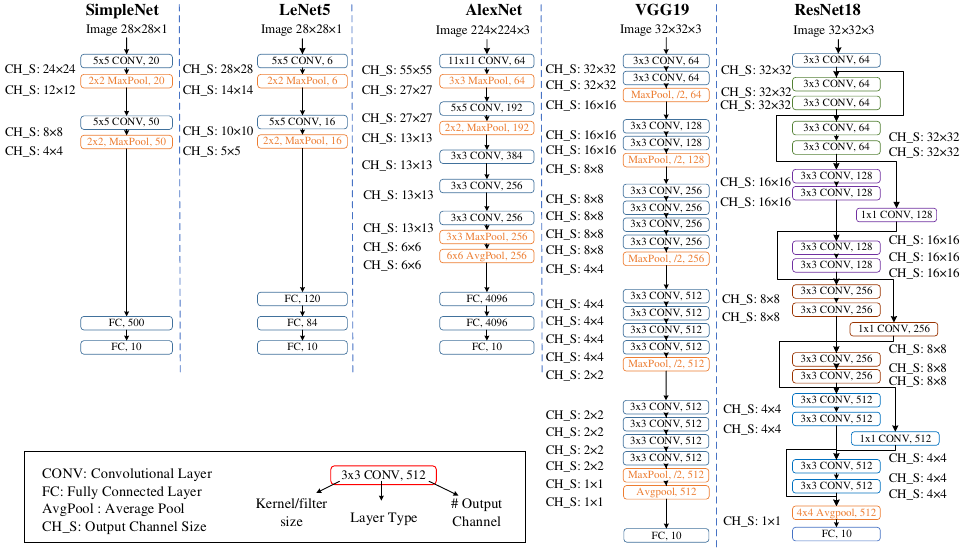}
	\caption{Different convolutional neural network architectures for image classification.}
	\label{fig5-CNN-Models}
\end{figure*}

To inject faults and analyze the vulnerability of the CNNs, first, the well-known \textit{SimpleNet}, \textit{LeNet5}, \textit{AlexNet}, \textit{ResNet18}, and \textit{VGG19} neural networks, whose architectures are shown in Fig.~\ref{fig5-CNN-Models} are trained using the \textit{PyTorch}~\cite{Paszke2019pytorch} framework for the \textit{MNIST}~\cite{lecun1998gradient} and~\textit{CIFAR10}~\cite{Krizhevsky09learningmultiple} datasets based on the parameters in Table~\ref{table1-relates-work}. \textit{LeNet5} and \textit{SimpleNet} are used to classify the \textit{MNIST} dataset and \textit{AlexNet}, \textit{ResNet18}, and \textit{VGG19} are used to classify the \textit{CIFAR10} dataset. \textit{MNIST} and \textit{CIFAR10} contain 60000 images for the training set and 10000 images for the test set. Top-1 accuracy is used as an evaluation parameter. Top-1 accuracy represents the percentage of test data instances correctly classified by the algorithm, with the highest predicted class being the same as the true class. We have developed our fault injection framework in Python using the \textit{PyTorch} framework. All simulations are done on an \textit{Intel Core i7, 3.2 GHz} processor with an \textit{NVIDIA GeForce GTX 2080Ti} GPU.

After training the CNN models, the trained model is placed in the inference phase, and fault injection is done in its parameters. Then again, for each fault injected, the top-1 accuracy of the test dataset for the defective network is calculated.
\begin{table*}[t]
	\centering
	\caption{Convolutional neural network specification for image classification.}
	\label{table1-relates-work}
	\begin{adjustbox}{max width=0.90\textwidth}
		\begin{tblr}{
				width = \linewidth,
				colspec = {Q[221]Q[160]Q[119]Q[137]Q[156]Q[127]},
				hline{1,17} = {-}{0.08em},
				hline{2,5,8,14,16} = {-}{},
			}
			Model                    & SimpleNet [41] & LeNet5 [2] & AlexNet [42] & ResNet18 [43] & VGG19 [3]  \\
			DataSet                  & MNIST          & MNIST      & Cifar10\_cts & Cifar10       & Cifar10    \\
			Image\_size              & 1×28×28        & 1×28×28    & 3×224×224    & 3×32×32       & 3×32×32    \\
			\# of Classes            & 10             & 10         & 10           & 10            & 10         \\
			Total Size (MB)          & 1.76           & 0.3        & 226.55       & 53.89         & 83.64      \\
			Total Parameters         & 431,080        & 44,426     & 57,044,810   & 11,173,962    & 20,040,522 \\
			Parameter size (MB)      & 1.64           & 0.17       & 217,61       & 42.63         & 76.45      \\
			\# of CONV Layer         & 2              & 2          & 5            & 20            & 16         \\
			Filter size              & 5              & 5          & 3,11,5       & 1,3           & 3          \\
			\# of in channel         & 1,20           & 1,6        & 3,256        & 3-512         & 3-512      \\
			\# of Filters [min, max] & {[}20,50]      & {[}6,16]   & {[}3,256]    & {[}64,512]    & {[}64,512] \\
			Stride                   & 1              & 1          & 1,2,4        & 1,2           & 1,2        \\
			\# of Parameters         & 25,570         & 2,572      & 2,469,696    & 11,168,832    & 20,035,392 \\
			\# of FC Layers          & 2              & 3          & 3            & 1             & 1          \\
			\# of Parameters         & 405,510        & 41,854     & 54,575,114   & 5,130         & 5,130      \\
			Accuracy                 & 99.42\%        & 99.26\%    & 86.82\%      & 89.42\%       & 88.40\%    
		\end{tblr}
	\end{adjustbox}
\end{table*}
\subsection{Fault Injection in Software-Level of Inference Phase} 
In this section, both single and multiple faults are injected at the software-level of CNN models in three phases to investigate the vulnerability of CNN model to different values of faults. In the first phase, the effects of single faults in each layer are investigated. In the second phase, the effect of variance in the weight of each layer is analyzed. In the third phase, we analyze behavioral simulation of stuck-at faults by injecting maximum and minimum values, to investigate the vulnerability of CNN to SaH and SaL faults. 
\subsubsection{Effect of each fault in each layer}
\begin{figure*}[t]
	\subfloat[]{\label{fig6-a-simpleNet}
		\includegraphics[width=0.33\linewidth]{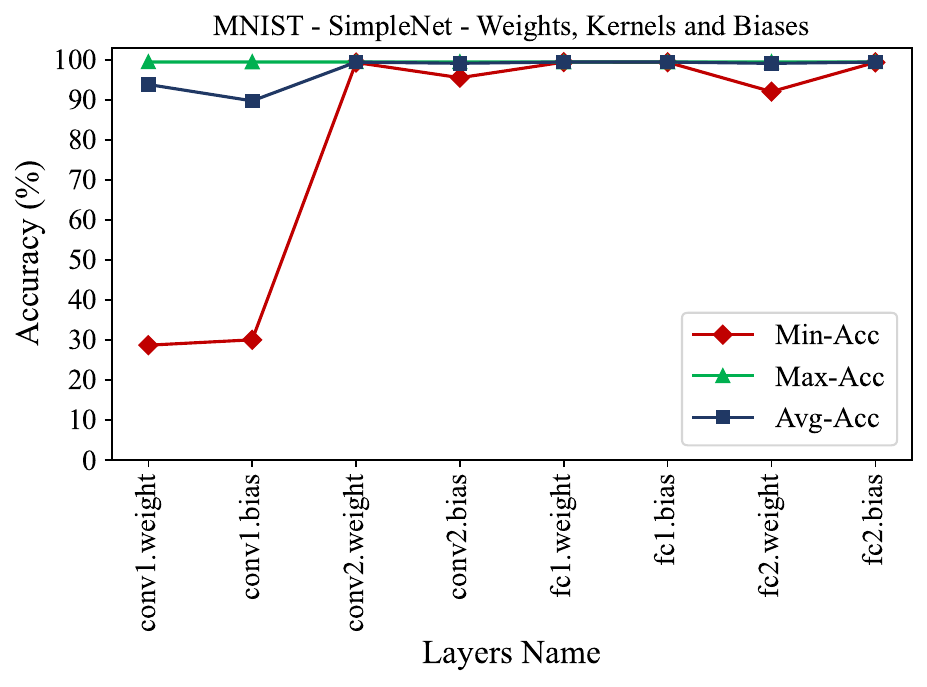}
	}
	\subfloat[]{\label{fig6-b-Lenet5}
		\includegraphics[width=0.33\linewidth]{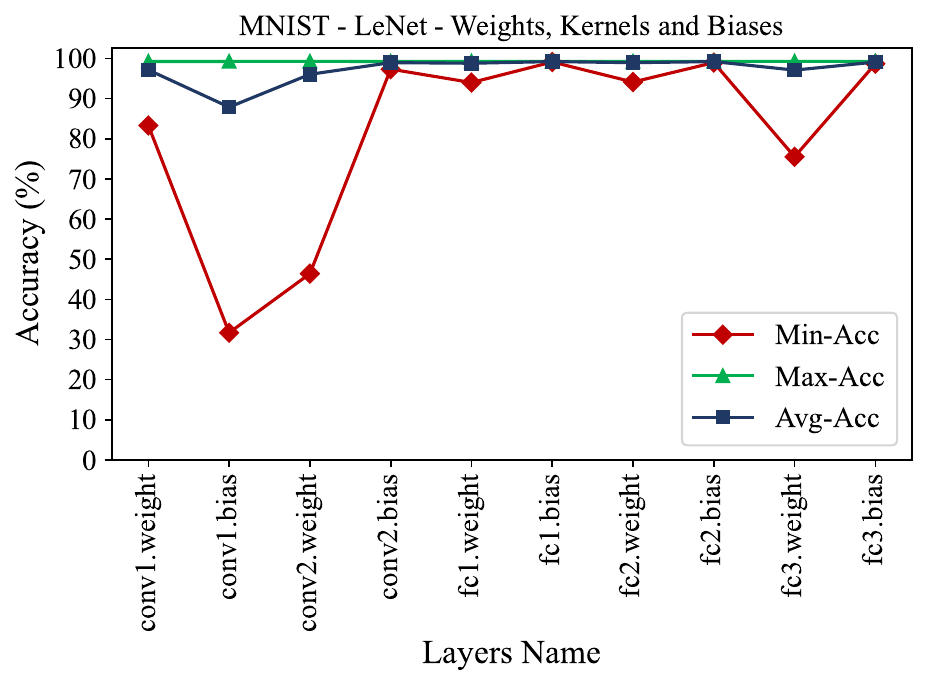}
	}
	\subfloat[]{\label{fig6-c-AlexNet}
		\includegraphics[width=0.33\linewidth]{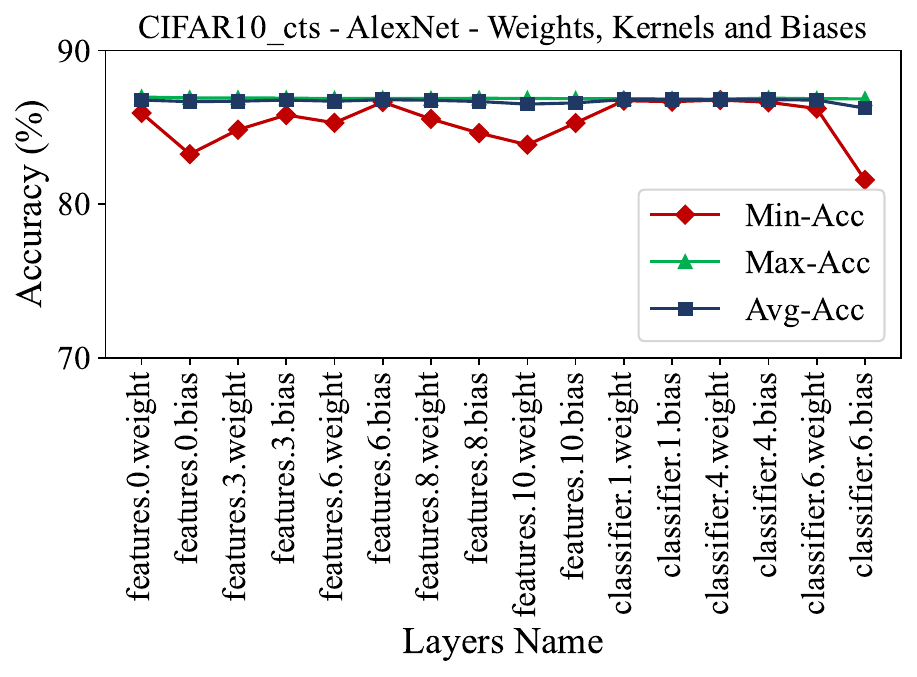}
	}
	\caption{The effect of single fault in different layers of several CNNs in (a) \textit{SimpleNet}, (b) \textit{LeNet5}, and (c) \textit{AlexNet}.}
	\label{fig6:multi1}	
\end{figure*}

\begin{figure*}[t]
	\subfloat[\label{fig7-a}]{
		\includegraphics[width=0.32\linewidth]{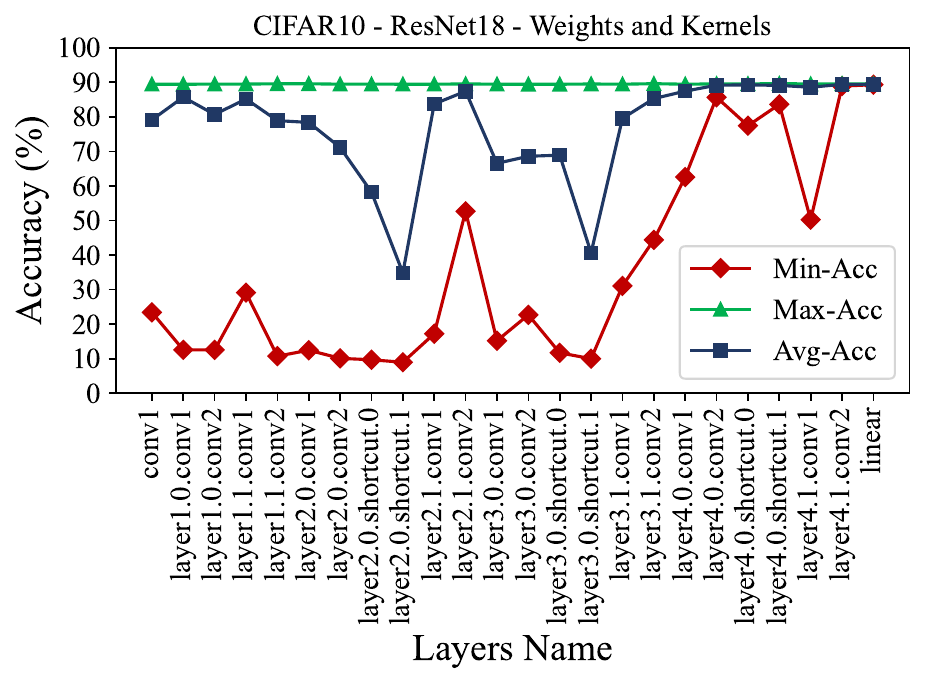}
	}
	\subfloat[\label{fig7-b}]{
		\includegraphics[width=0.32\linewidth]{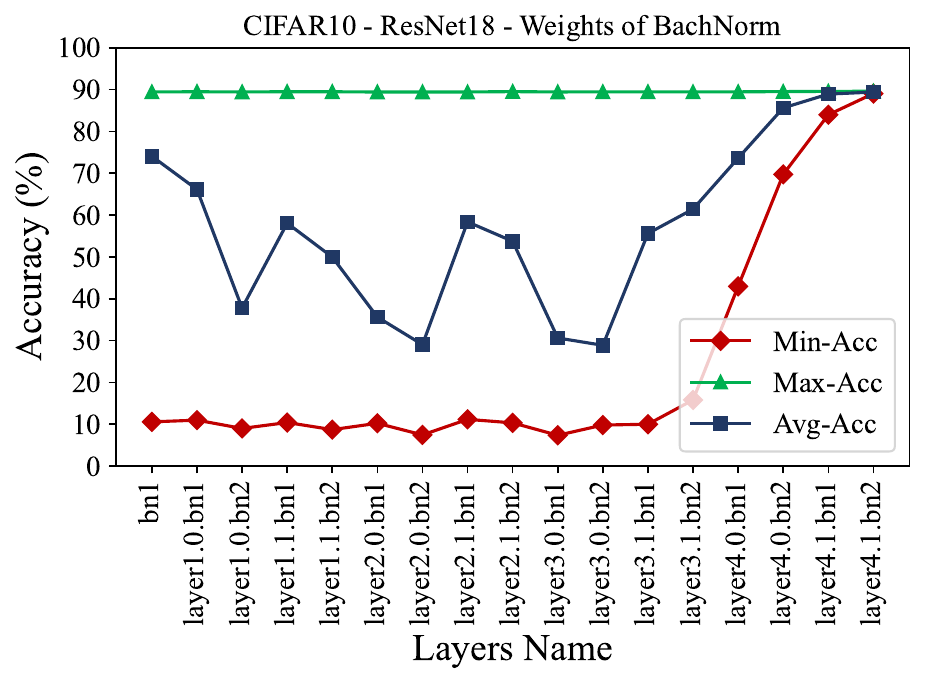}
	}
	\subfloat[\label{fig7-c}]{
		\includegraphics[width=0.32\linewidth]{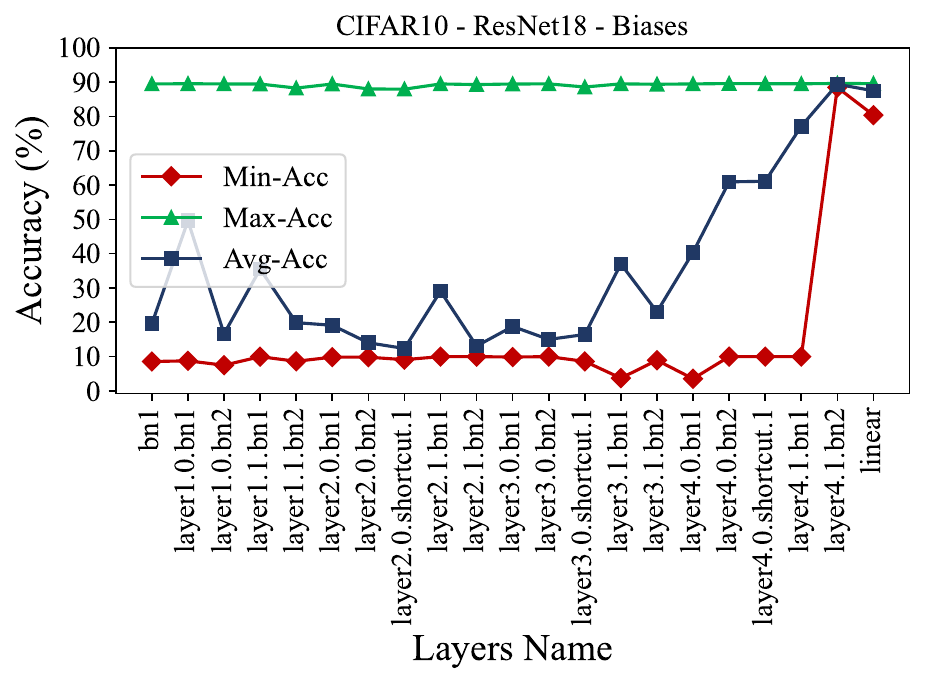}
	}
	\caption{The effect of single fault in different layers of several CNNs in \textit{ResNet18}. a) Filters of \textit{CONV}s and \textit{FC}, b) Weights of \textit{BachNorms} and c) Biases of \textit{BachNorms} and \textit{FC}.}
	\label{fig7:multi1}	
\end{figure*}

\begin{figure*}[!h]
	\centering
	\subfloat[\label{fig8-a}]{
		\includegraphics[width=0.35\linewidth]{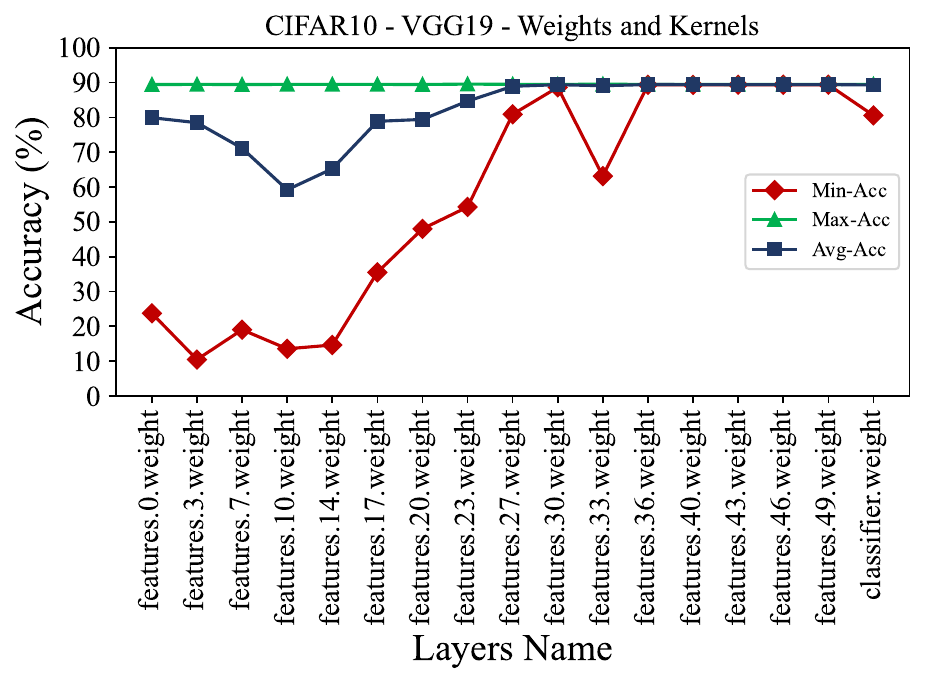}
	}
	\subfloat[\label{fig8-b}]{
		\includegraphics[width=0.35\linewidth]{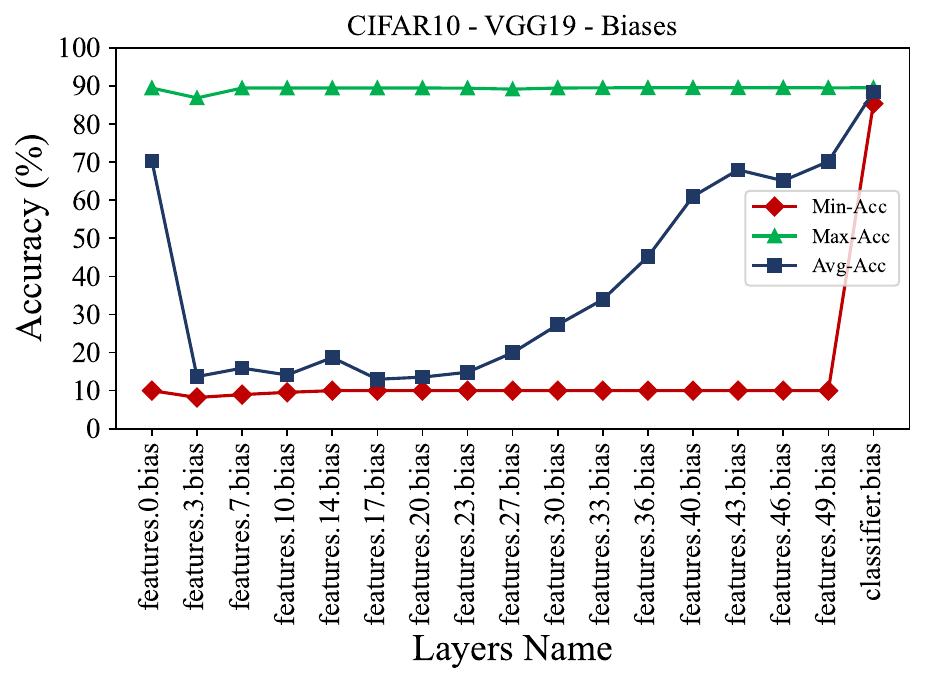}
	}
	
	\subfloat[\label{fig8-c}]{
		\includegraphics[width=0.35\linewidth]{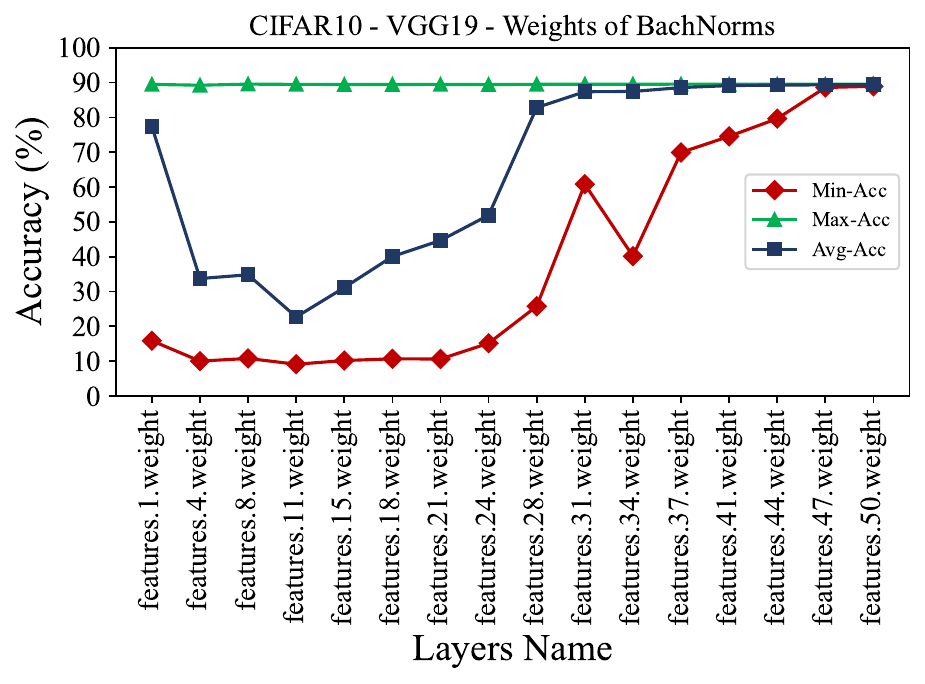}
	}
	\subfloat[]{
		\includegraphics[width=0.35\linewidth]{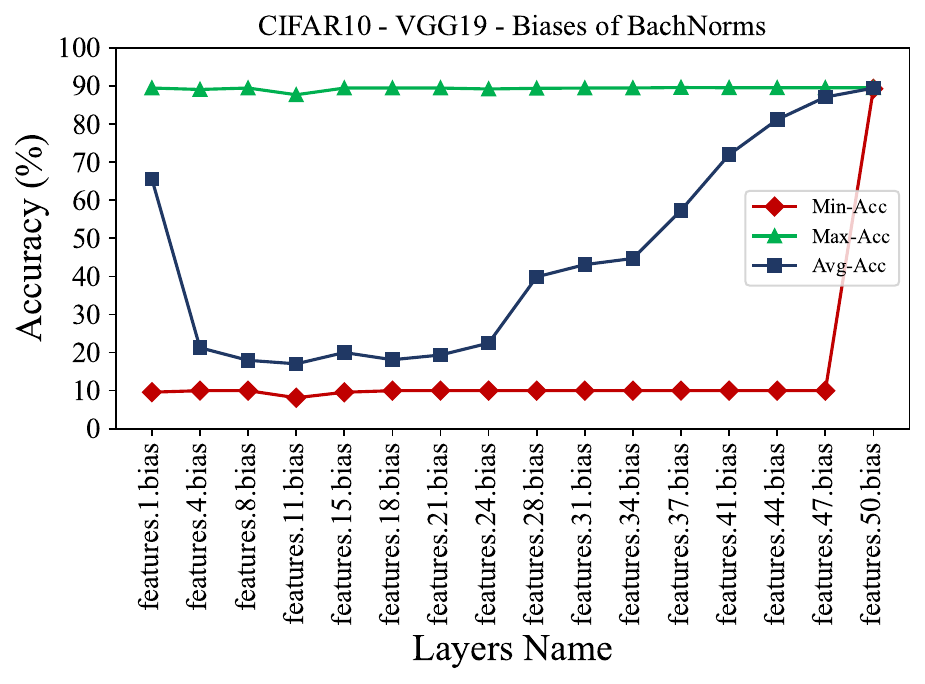}
	}
	
	\caption{The effect of single faults in different layers of several CNNs in VGG19. a) Filters of~\textit{CONVs} and \textit{FC}, b) Bias of \textit{CONVs} and \textit{FC}, c) Weights of\textit{BachNorms}, and d) Bias of \textit{BachNorms}.}
	\label{fig8:multi3}	
\end{figure*}

In the first phase, 10 million faults consisting of 1000 faults for 10000 input images are injected at random locations of weights, kernels, and biases of each layer at the algorithmic level to determine the sensitivity of each layer to faults. The maximum, minimum, and average top-1 classification accuracies of 1000 repetitions of the 10000 images of the test case for the different models are given in figures~\ref{fig6:multi1} through~\ref{fig8:multi3}.
\begin{table*}[t]
	\centering
	\caption{Accuracy of faulty models and number of experiments affected by faults in different CNN models in the sense of single faults.}
	\label{table2-accuracy-of-different-model}
	\begin{adjustbox}{max width=0.90\textwidth}
	\begin{tblr}{
			width = 0.9\linewidth,
			colspec = {Q[75]Q[123]Q[192]Q[137]Q[129]Q[127]Q[135]},
			column{even} = {c},
			column{3} = {c},
			column{5} = {c},
			column{7} = {c},
			vline{5} = {-}{0.08em},
			hline{1-2,7} = {-}{0.08em},
		}
		CNN Models & Baseline Accuracy (\%) & Average accuracy of all Faulty Layers (\%) & Accuracy Degradation (\%) & Degraded Accuracy (\%) & Improved Accuracy (\%) & Unchanged Accuracy (\%) \\
		SimpleNet  & 99.42                  & 97.41                                      & 2.01                      & 65.13                  & 7                      & 27.88                   \\
		LeNet5     & 99.26                  & 97.25                                      & 2.01                      & 72.6                   & 8.9                    & 18.5                    \\
		AlexNet    & 86.82                  & 86.61                                      & 0.21                      & 60.88                  & 20.69                  & 18.44                   \\
		ResNet18   & 89.41                  & 56.87                                      & 32.54                     & 80.81                  & 7.29                   & 11.9                    \\
		VGG19      & 88.4                   & 57.72                                      & 30.68                     & 72.77                  & 13.61                  & 13.62                   
	\end{tblr}
\end{adjustbox}
\end{table*}

By injecting faults into the learning parameters, the classification accuracy of CNNs is altered depending on the type and depth of layers and the location of faults. Interestingly, the effect of faults on CNNs is not always in the direction of degrading accuracy. Since the value of the learning parameter changes over an optimal local value in the training phase, it is probable that changing the parameter after the training phase improves classification accuracy, which  rarely happens in different experiments. For example, in the \textit{SimpleNet} model, fault injections in the Conv1 layer in 2\% of experiments improve the accuracy, while 80\% of experiments decrease the accuracy and in 18\% of experiments, no change in the accuracy of faulty models is observed. Table~\ref{table2-accuracy-of-different-model} contains two sections: The first shows the average accuracy, lesser accuracy, and equal accuracy of faulty layers when it comes to classification accuracy degradation. It also shows how many smaller, greater, and equal simulations were run with baseline accuracy.

The result of injecting faults into different layers of \textit{SimpleNet}, \textit{LeNet5}, and \textit{AlexNet} is shown in Fig.~\ref{fig6:multi1}. Injecting faults in weights of \textit{FC} layers and kernels of \textit{CONV} layers, and also learning parameters of \textit{BachNorm} for \textit{ResNet18} and \textit{VGG19} is done as shown in Fig.~\ref{fig7:multi1} and Fig.~\ref{fig8:multi3}, respectively. As shown, in the \textit{CONV} layer, occurring a fault in the value of kernels causes all the values of the corresponding output channel to be erroneous. This event in the first layers causes a large number of the output value of subsequent layers to change, as the error is propagated to the subsequent layers and increases the probability of final error in the output of CNN. However, the occurrence of a fault in the last \textit{CONV} layers has less effect on the final accuracy of the network than in FC layers. As we go into the depth of the neural network, the sensitivity of CNN to faults in the \textit{CONV} layers decreases, and it increases the accuracy in the \textit{DENSE} or \textit{FC} faulty layers. It is observed that as the number of \textit{FC} layers increases, the vulnerability of neural networks to faults increases as well. On the other hand, some of the faults are masked in these layers. For example, in the \textit{AlexNe}t model, in Fig.~\ref{fig7-c} with 3 \textit{FC} layers and 54 million parameters in these layers, there is just 0.21\% accuracy degradation compared to the baseline model. Shortcut connections in the \textit{ResNet18} model cause more errors propagation to the next block of \textit{ResNet18} or other layers (Fig.~\ref{fig7:multi1}).

As reported in Table~\ref{table2-accuracy-of-different-model}, \textit{ResNet18} and \textit{VGG19} networks, with a significantly larger number of parameters and \textit{CONV} layers than \textit{AlexNet} and just one \textit{FC} layer, are more vulnerable to faults. Shortcut connections in \textit{ResNet} cause error propagation to subsequent layers. Changing the learning parameter of the \textit{BachNorm} layer significantly reduces the classification accuracy in \textit{ResNet18} and \textit{VGG19}. 

\begin{figure*}[t]
	\centering
	\subfloat[\label{fig9-a}]{
		\includegraphics[width=0.80\linewidth]{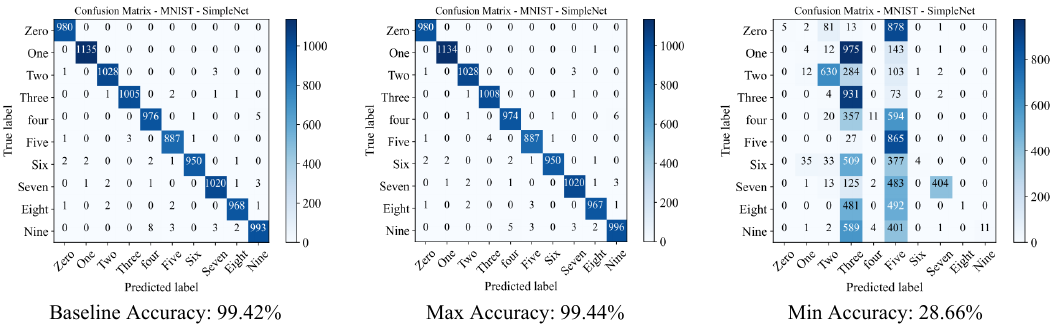}
	}
	
	\subfloat[\label{fig9-b}]{
		\includegraphics[width=0.80\linewidth]{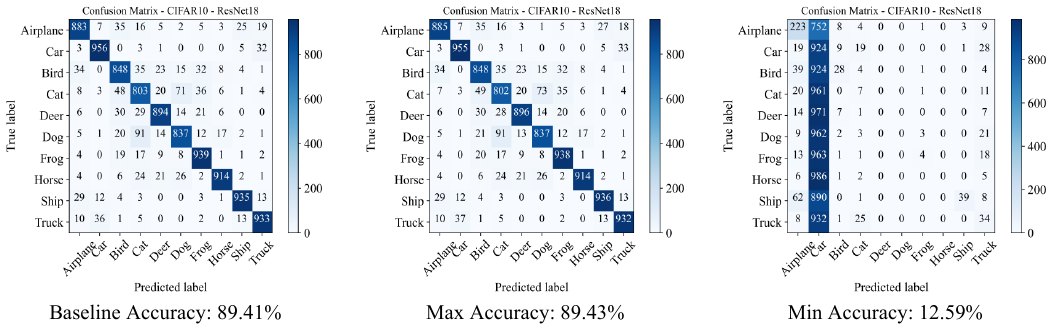}
	}

	\subfloat[\label{fig9-c}]{
		\includegraphics[width=0.80\linewidth]{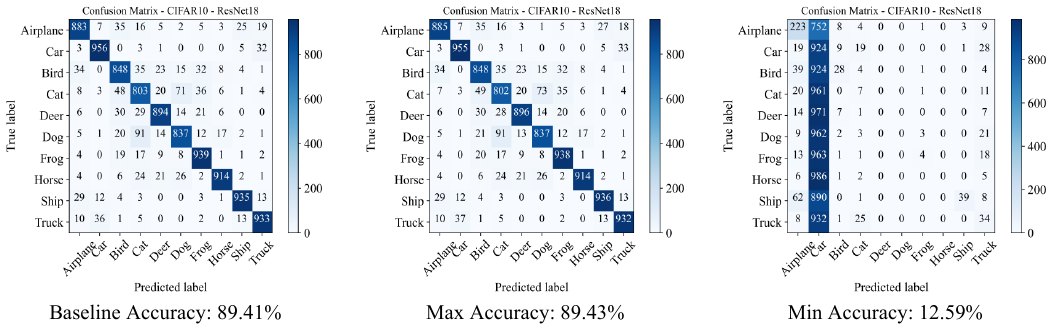}
	}
	\caption{Confusion matrix of baseline accuracy compared with minimum and maximum accuracy of faulty CNN. Classification of (a) MNIST dataset with \textit{SimpleNet} model, (b) \textit{MNIST} dataset with \textit{LeNet5}, and (c) \textit{Cifar10} dataset with \textit{ResNet18}.}
	\label{fig9:multi3}	
\end{figure*}

As mentioned, any error in the value of a learning parameter of the CNN changes the network's output. It is probable in some scenarios that some input samples correctly predicted in the baseline model are predicted incorrectly in the faulty model, or conversely, the samples that were wrongly predicted in the baseline model are correctly predicted in the presence of faults. In this regard, to more accurately evaluate the performance of the neural network in the presence of faults and fault-free conditions, the confusion matrix has been used, which specifies the predicted output and the actual output for the test datasets. Each row specifies the actual label of the class, and each column specifies the number of predicted examples in that class. The value of the main diameter represents the correctly predicted samples, and the value outside the main diameter represents the incorrectly predicted classes. The confusion matrix of faults for the baseline model, maximum accuracy, and minimum accuracy in the first layer of the three models are shown in Fig.~\ref{fig9:multi3}.

\subsubsection{Effect of variance in weights}
\begin{figure*}[t]
	\centering
	\subfloat[\label{fig10-a}]{
		\includegraphics[width=0.90\linewidth]{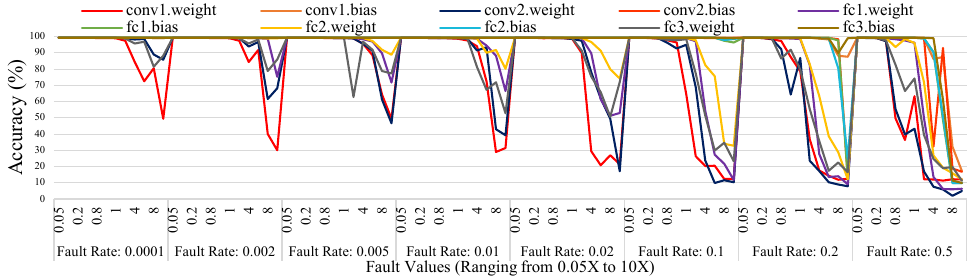}
	}
	
	\subfloat[\label{fig10-b}]{
		\includegraphics[width=0.90\linewidth]{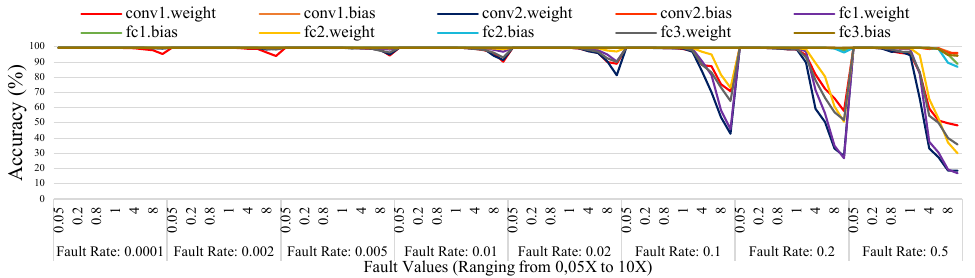}
	}
	
	\caption{CNN accuracy under different number of faults injected to weights and kernels by different value in \textit{LeNet5}. (a) Minimum accuracy and, (b) Average accuracy.}
	\label{fig10:multi3}
\end{figure*}

\begin{figure*}[!t]
	\centering
	\subfloat[\label{fig11-a}]{
		\includegraphics[width=0.90\linewidth]{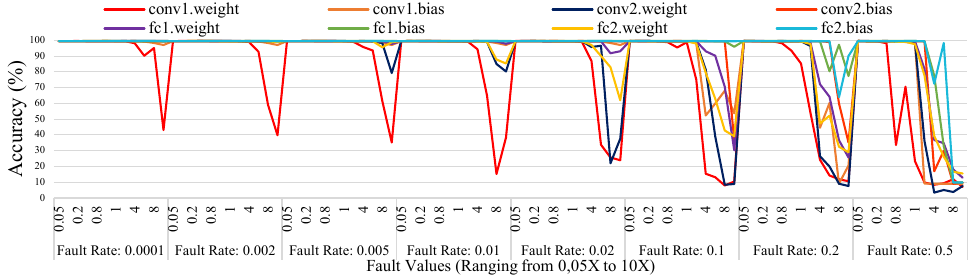}
	}
	
	\subfloat[\label{fig11-b}]{
		\includegraphics[width=0.90\linewidth]{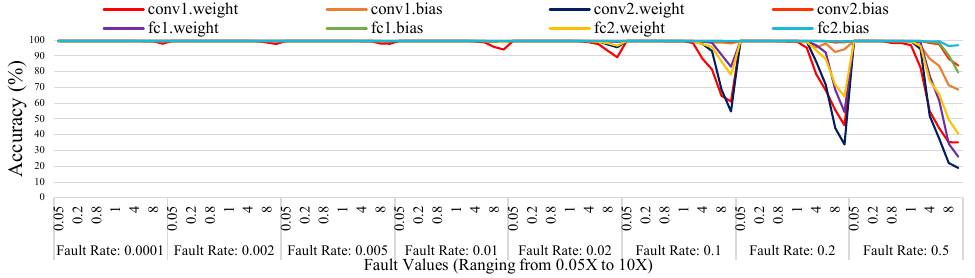}
	}
	
	\caption{Accuracy of different number of faults injected to weight or kernel by different value in \textit{SimpleNet}. (a) Minimum accuracy and, (b) Average accuracy.}
	\label{fig11:multi3}	
\end{figure*}

In the second phase of software-level fault injection in CNN models, the focus is on stuck-at value faults (SaV) in such a way that different numbers of weights, kernels, and biases are stuck-at a specific value. To this aim, values of weights or kernels and bias are randomly changed at different times from $0.05X$ to $10X$ of the golden value. It means that $X$ times the value has been added to it. The accuracy of the different numbers of elements and different ranges of changes for \textit{SimpleNet} and \textit{LeNet5} are shown in Fig.~\ref{fig10:multi3} and Fig.~\ref{fig11:multi3}, respectively. As shown, by increasing the number of faults and larger deviations from the golden value, classification accuracy decreases significantly. The first layers have great vulnerability to faults, even in small number of faults. By increasing the fault rate, successive layers will expose their vulnerability to faults. As shown in these figures, by 20\% faults and data change range of 80\%, the accuracy degrades to 7\%-20\%, which is extremely less than a random and untrained network.

\subsubsection{Effect of multiple SaH and SaL faults in each layer}
In the third step, to study the sensitivity to multiple SaH and SaL faults, more than one fault with the value of the maximum and minimum amounts have been injected into the weight, kernel, and bias matrices. The maximum and minimum values indicate SaH resistance and SaL resistance, respectively. For this reason, a portion of the learning parameter has been modified at random positions in each layer of the \textit{LeNet5} and \textit{SimpleNet} models to either the maximum or minimum value, or a combination of both.

\begin{figure*}[t]
	\centering
	\subfloat[\label{fig12-a}]{
		\includegraphics[width=0.45\linewidth]{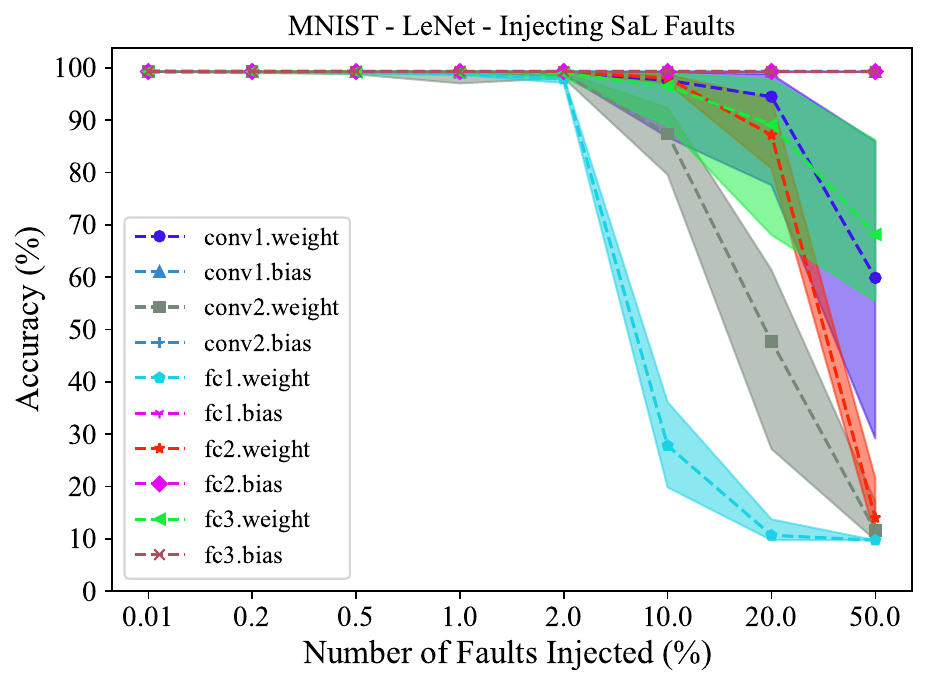}
	}
	\subfloat[\label{fig12-b}]{
		\includegraphics[width=0.45\linewidth]{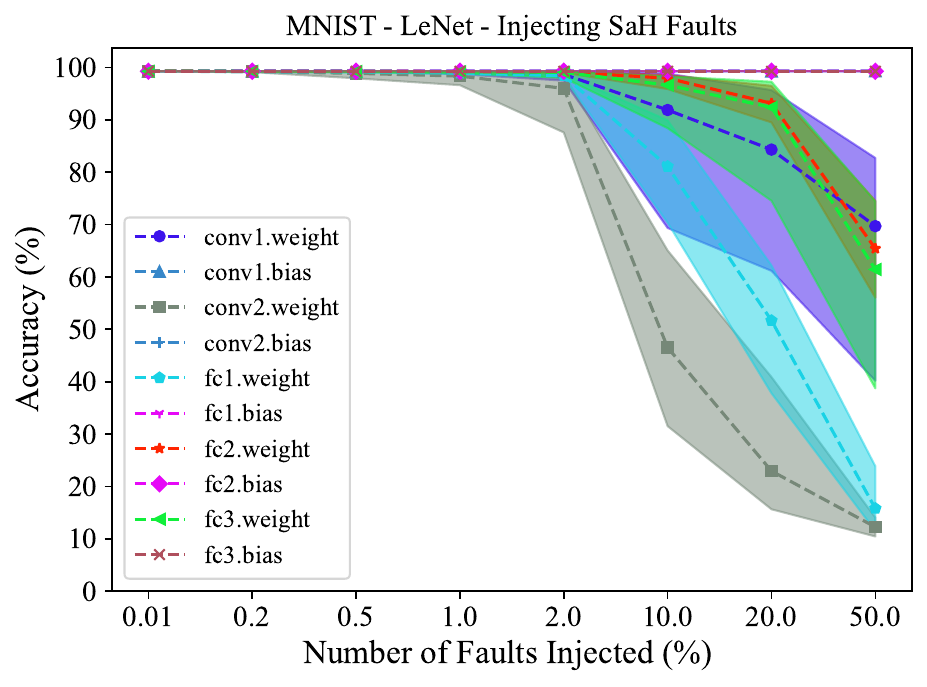}
	}
	\caption{Injection of faults in different layers of \textit{LeNet5}, (a) SaL and (b) SaH}
	\label{fig12:multi3}
\end{figure*}

\begin{figure*}[!t]
	\centering
	\subfloat[\label{fig13-a}]{
		\includegraphics[width=0.45\linewidth]{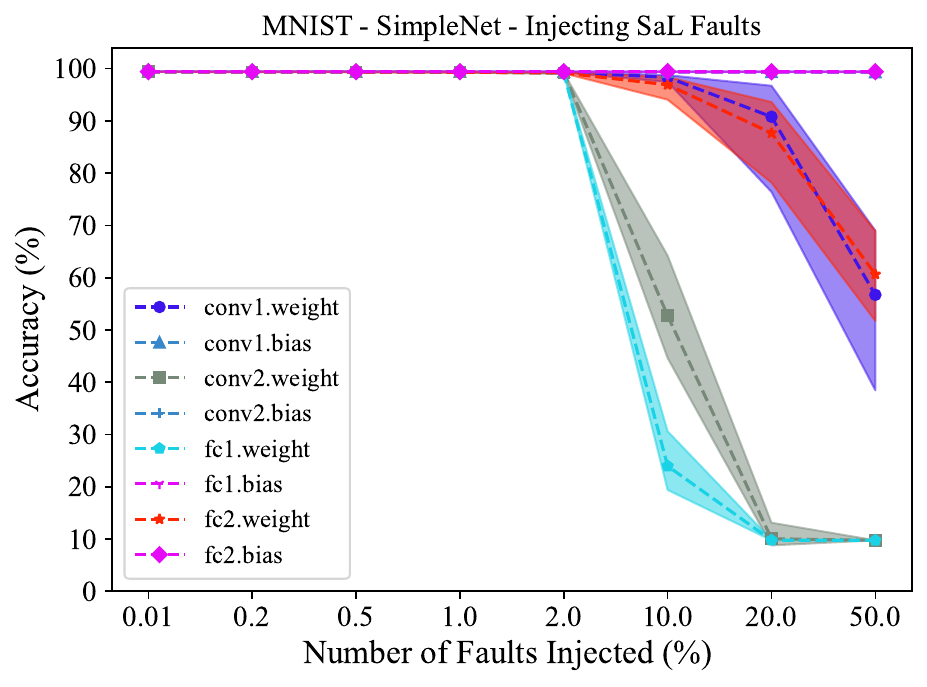}
	}
	\subfloat[\label{fig13-b}]{
		\includegraphics[width=0.45\linewidth]{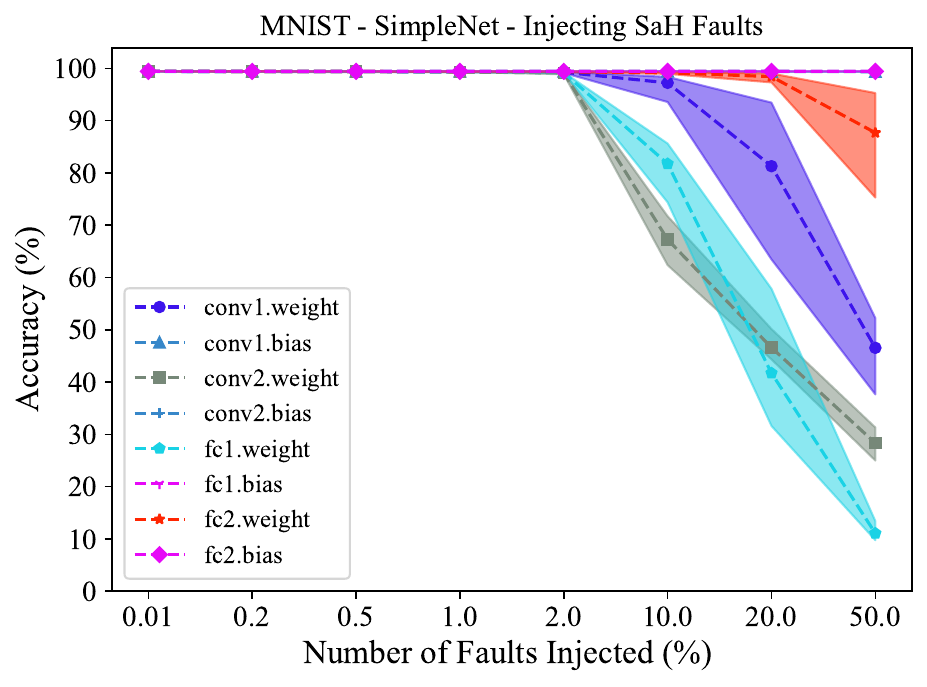}
	}
	\caption{Injection of faults in different layers of \textit{SimpleNet}, (a) SaL and (b) SaH}
	\label{fig13:multi3}
\end{figure*}

\begin{figure*}[!t]
	\centering
	\subfloat[\label{fig14-a}]{
		\includegraphics[width=0.45\linewidth]{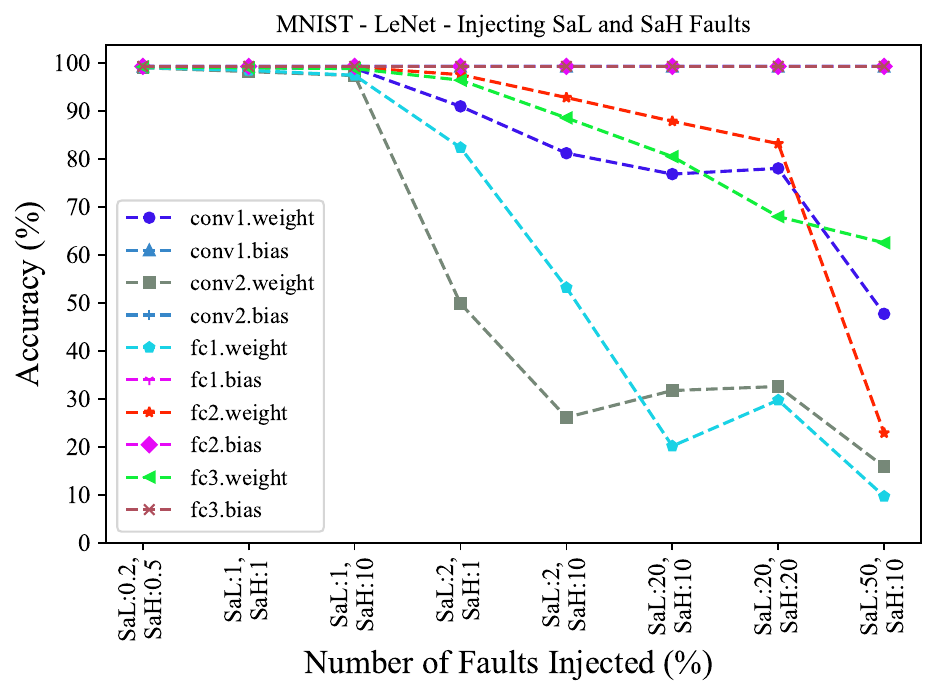}
	}
	\subfloat[\label{fig14-b}]{
		\includegraphics[width=0.45\linewidth]{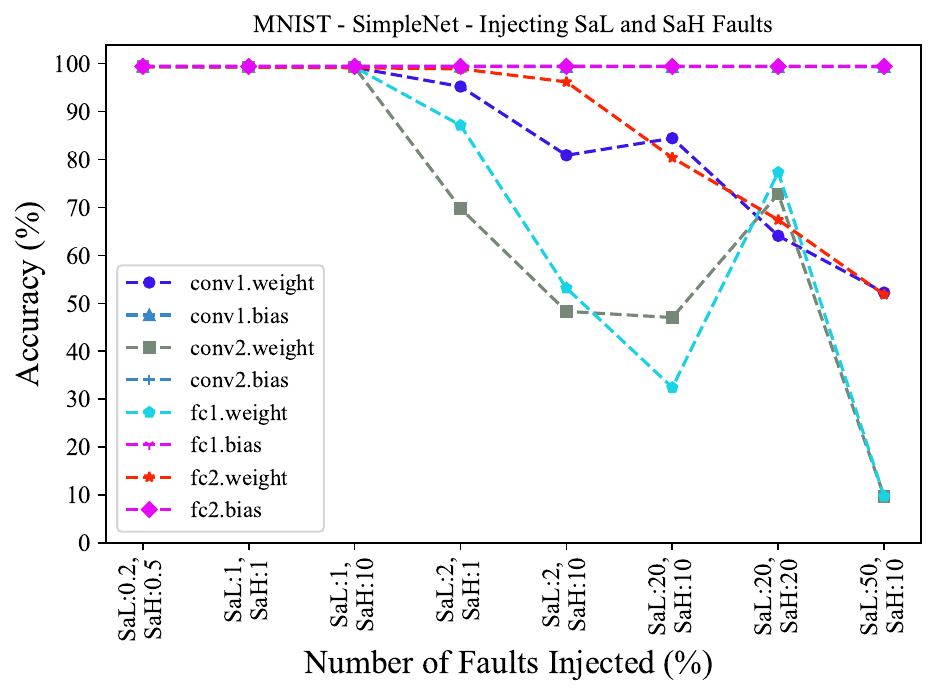}
	}
	\caption{Injection of combination of maximum and minimum faults in different layers of (a) \textit{LeNet5} and (b) \textit{SimpleNet}}
	\label{fig14:multi3}
\end{figure*}
The results of injecting SaL and SaH faults in different layers of \textit{LeNet5} and \textit{SimpleNet} models are shown in Fig.~\ref{fig12:multi3} and Fig.~\ref{fig13:multi3}, respectively. Fig.~\ref{fig12-a} and~\ref{fig13-a} show the accuracy of networks resulting from injection of the SaL faults, and Fig.~\ref{fig12-b} and Fig.~\ref{fig13-b} show the results of injecting SaH faults. In these figures, each line shows the accuracy of injecting faults in different iterations of experiments in its corresponding layers while the other layers are fault-free. In each fault rate, maximum, minimum, and average fault injection experiments are denoted by a specific line. As shown, by increasing the number of faulty cells, the accuracy of the neural network decreases drastically. In addition, the effect of SaL faults on accuracy degradation is significantly higher than that of SaH faults. Fig.~\ref{fig14:multi3} shows that the results of the combination of SaL and SaH faults are almost similar to the results of Fig.~\ref{fig12-a} and~\ref{fig13-a}, which indicates the greater vulnerability of the neural network to SaL faults.

\begin{figure}[!t]
	\centering
	\subfloat[]{\label{fig15-a}
		\includegraphics[width=0.9\linewidth]{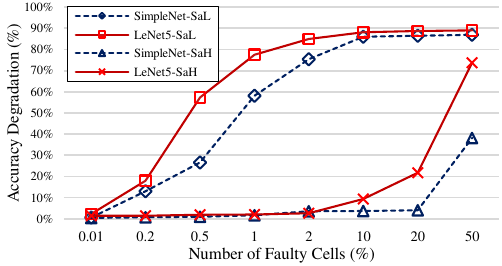}
	}
	
	\subfloat[]{\label{fig15-b}
		\includegraphics[width=0.9\linewidth]{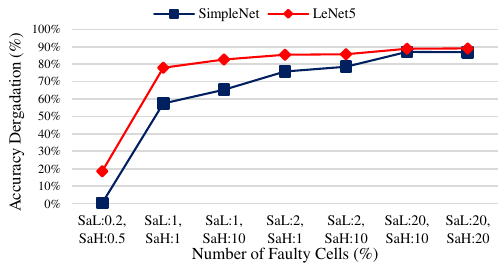}
	}
	\caption{Accuracy degradation over injecting faults in  \textit{LeNet5} and \textit{SimpleNet}, (a) just SaL or SaH and (b) both SaL and SaH.}
	\label{fig15:multi3}
\end{figure}

\begin{figure}[t]
	\centering
	\subfloat[]{\label{fig16-a}
		\includegraphics[width=0.70\linewidth]{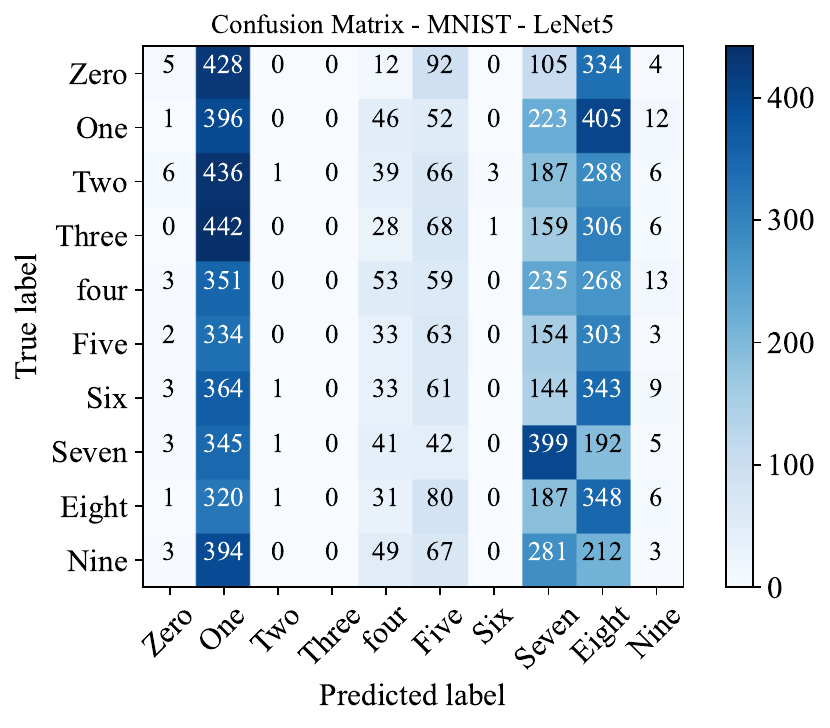}
	}
	
	\subfloat[]{\label{fig16-b}
		\includegraphics[width=0.70\linewidth]{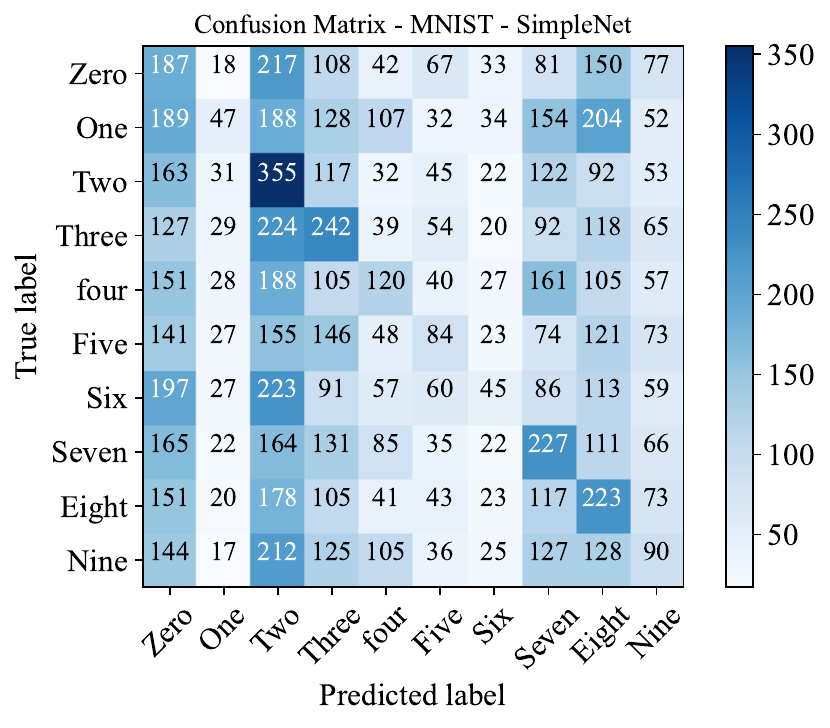}
	}
	\caption{Confusion matrix by injecting 10\% SaH and 20\% SaL for classification of MNIST with (a) \textit{LeNet5} and, (b) \textit{SimpleNet}}
	\label{fig16:multi3}
\end{figure}
\subsection{Fault Injection in Hardware-Level of Inference Phase}
Fault injection in the processing platform of the ReRAM-based accelerator is done by changing the MemTorch simulator~\cite{lammie2020memtorch} which models ReRAM-based accelerators. This simulator maps the network parameters as the conductance of the ReRAM crossbar array as~\eqref{eq4} and inputs are scaled to [-0.3,+0.3],
\begin{equation}
	\label{eq4}
	g[i,j]=\frac{(g_{max}-g_{min})(w[i,j]-w_{min} )}{(|w|_{max}-w_{min})+g_{off}}
\end{equation}
where $w$ is the weight and kernel parameters and $g$ is the conductance of ReRAM cells. $|w|_{max}$ and  $w_{min}$ are the maximum of absolute value and minimum of learning parameters, respectively. The positive values are mapped to positive ReRAM crossbar arrays and the negative values are mapped to the conductance of negative crossbar arrays.

After mapping the learning parameters of CNN to the conductance of ReRAM crossbar cells, SaH and SaL faults are injected into random cells of the crossbar. Then, the faulty CNN is tested on the test dataset, and the top-1 classification accuracy is calculated. The simulation has been performed for different numbers of SaH, SaL, and a combination of both SaH and SaL faults. SaL=10 means that 10\% of ReRAM cells are stuck to the low resistance state, and SaH=0.01 means that 0.01\% of ReRAM cells are stuck to the high resistance state. The results of fault injection into \textit{LeNet5} and \textit{SimpleNet} CNN models mapped to a ReRAM array are shown in Fig.~\ref{fig15:multi3}. In Fig.~\ref{fig15-a}, the dashed line shows the accuracy degradation to the baseline model in the \textit{SimpleNet}, and the solid line shows the accuracy degradation due to SaL and SaH in \textit{LeNet5} model faults. Fig.~\ref{fig15-b} shows accuracy degradation due to the occurrence of both SaL and SaH faults. 

As the number of faulty ReRAM cells increases, the CNN model will have more learning parameters altered, and therefore the error of the CNNs will increase. When comparing Fig.~\ref{fig15-a} and Fig.~\ref{fig15-b}, it is clear that the degree of accuracy reduction is dependent on SaL faults, indicating that the neural network is more vulnerable to SaL resistance faults than SaH resistance faults. A recent study of ReRAM cells reported that 9.04\% and 1.75\% of ReRAM cells are stuck-at high and stuck-at low resistance states, respectively~\cite{chen2014rram} indicating significantly higher occurrence probability of SaH. With this aforementioned assumption and according to the data in Fig.~\ref{fig15:multi3}, this number of faults in ReRAM crossbar cells can reduce the accuracy of the CNNs between 50\% and 70\% in the presence of SaL and about 10\% in the presence of SaH. Although the percentage of SaH faults is higher than SaL faults, the vulnerability of SaL is much higher, so tolerating SaL faults should be prioritized.

Fig.~\ref{fig16:multi3} depicts a confusion matrix generated by running \textit{LeNet5} and \textit{SimpleNet} on a ReRAM-based CNN accelerator when the crossbar array has 20\% SaL and 10\% SaH. In these matrices, most of the images are predicted in the wrong class. The classification accuracy of faulty networks are about 10.23\% and 10\% in \textit{LeNet5} and \textit{SimpleNet}, respectively.

\section{Summery and Discussion }\label{Summery and Discussion}
Starting from the behavioral model of CNN under study, the following observations were obtained by injecting different fault models at the algorithm level and in the conductivity of ReRAM crossbar cells. According to these findings, efficient methods for designing a fault-tolerant architecture in ReRAM-based CNN accelerators can be proposed:
\begin{enumerate}
	\item{Different parameters have a different impact on classification accuracy. These parameters contain the topology of CNNs, the types and the number of each layer, the distance of each faulty layer from the output layer, and the value and types of faults.}
	\item{The different parameters have different vulnerabilities. For example, bias parameters in \textit{CONV} layers are more immune to faults, while weights in both \textit{CONV} and \textit{DENSE} layers have a high impact on accuracy.}
	\item{As the deviation of the faulty value from the original value increases, the classification error increases more, and by increasing the number of faulty cells, the error propagates more in layers.}
	\item{Compared to other layers such as \textit{DENSE} layers, convolutional layers are more vulnerable to faults. As we go deeper into the CNNs, the vulnerability of the convolutional layers decreases.}
	\item{The learning parameters of the \textit{BachNorm} layers are extremely vulnerable to the occurrence of faults. The classification accuracy of CNNs decreases significantly by changing the weights and biases of these layers.}
	\item{Occurrence of shortcut connections in faulty layers propagates error in multiple layers and has high a impact on degradation of classification accuracy.}
	\item{The number of \textit{DENSE} layers affect the outputs of the faulty network. As the number of \textit{DENSE} layers increases, the propagated errors in \textit{CONV} layers decrease in \textit{DENSE} layers. Moreover, models with just one \textit{DENSE} layer such as \textit{VGG19} and \textit{ResNet18} are more vulnerable to faults.}
	\item{The occurrence of faults in the first layers of the convolutional layer causes error propagation in the networks and further reduces the accuracy of the CNNs.}
	\item{There is a large variation in vulnerability to SaH and SaL faults. On the other hand, the probability of SaH and SaL is also different. Although the percentage of SaL cells in the crossbar array is lower than that of SaH, the vulnerability of neural networks to SaL faults is higher.} 
\end{enumerate}

It should be noted that as the number of layers increases, the computation time of exploration for different fault values and different numbers of faults in each layer will increase polynomially. Our current focus on simpler models allows us to perform more comprehensive exploration within the constraints of the study. However, a more extensive study using variable CNNs still appears to show the same trend. In addition, by utilizing time and resource efficient fault injection methods, we would use more complex CNNs as part of our future work.

\section{Related Work}\label{Related Work}
Fault occurrence during neural network execution can cause errors in the output of the networks. For this reason, previous work has investigated and analyzed the types of faults and errors in these networks. The reliability of CNN accelerators such as GPU, FPGA, and ASIC against transient faults is analyzed in~\cite{ibrahim2020soft}. There is another gap in analyzing the type of faults in analog computational ReRAM. By injecting faults into different layers and parameters of the network, it is possible to understand the vulnerability of neural networks to faults. So far, several fault-injection tools have been proposed for different DNN models running on different accelerator platforms (GPU, TPU, and ASIC). Transistor-level fault simulation is only done at the neuron level of spiking neural network (SNN)~\cite{el2019self} or a small size network~\cite{temam2012defect}. These models run on specific spiking-based hardware, which is for running just small and specific SNN or DNN models. In general, large-scale fault injection simulations require a fault model with a high level of abstraction. It also enables a higher level of reliability analysis that is independent of hardware implementation characteristics. For this purpose, experiments for fault injection have been performed in several aspects using the following:
\begin{enumerate}
 \item{the behavioral level of abstraction of faults in synapse or neuron of SNN~\cite{vatajelu2019special, el2020spiking};}
 \item{permanent and transient bit-flip in the data path and memory to implement the CNN models~\cite{li2017understanding, mahmoud2020pytorchfi, reagen2018ares, neggaz2019cnns, hari2017sassifi, tsai2021nvbitfi, chen2020tensorfi};}
 \item{stuck-at fault at gate-level for TPU~\cite{zhang2019fault}, in the resolution of activation in FPGA~\cite{gambardella2019efficient}, and conductance of ReRAM crossbar cells [29].}
 \end{enumerate}
 The introduced tools deal with non-ReRAM accelerators or are for storage-based ReRAMs. The hardware methods consider fault injections for some specific small DNN models while the design of a ReRAM-based PIM accelerator requires novel methods able to capture thesystem vulnerability at a broader system level. 

\textit{TensorFi}~\cite{chen2020tensorfi} is a fault injection tool based on the \textit{tensorflow} framework, and \textit{pytorchFi}~\cite{mahmoud2020pytorchfi} is based on \textit{PyTorch} framework in algorithm-level proposed for fault injection in the neural network. TensorFi injects faults in the data path of neural networks and \textit{PyTorchFi} changes the value of neurons and weights in specific locations. They are merely used to inject faults through random data changes without analyzing the vulnerability of different CNN models or taking the hardware platform into account. \textit{SASSIFI}~\cite{hari2017sassifi} and \textit{NVBITFI}~\cite{tsai2021nvbitfi} are two tools for fault injection at the instruction-level of GPU can also be used for CNNs. 

A simulation is proposed in~\cite{li2017understanding} to inject random faults into the data path of \textit{AlexNet}, \textit{CaffeNet}, \textit{NiN}, and \textit{ConvNet} for several data types of floating-point and fixed-point with different numbers of bits. Then, hardware and software methods are proposed for the \textit{Eyeriss} NN accelerator~\cite{7551407_eyeriss} to reduce silent data corruption (SDC) errors. In a typical application, SDC is the result of a failure when the output of the application is different from the expected output. The bit-by-bit comparison of the generated output and the original one is done to detect errors. However, the bit-by-bit comparison is meaningless in neural networks. Therefore, in this article, a new definition is presented for SDC which means that the predicted output class label is different from the desired input.

It should be noted that the aforementioned tools, frameworks, and methods (described in detail in Table~\ref{table3-related-work}) are either intended to deal with software-level fault injection or are exclusively for non-ReRAM accelerators. Consequently, in this paper, the source of faults, errors, and reduction of accuracy in ReRAM-based PIM accelerators has been investigated and a detailed analysis of the effects of faults at different places of the neural network has been carried out based on the characteristics of the ReRAM accelerator.

\begin{table*}[t]
	\centering
	\caption{Summary of existing fault injection scheme.}
	\label{table3-related-work}
\begin{adjustbox}{max width=0.90\textwidth}
	\begin{tblr}{
		colspec = {Q[90]Q[87]Q[108]Q[148]Q[142]Q[98]Q[67]Q[104]Q[87]},
		row{10} = {c},
		row{15} = {c},
		row{17} = {c},
		row{19} = {c},
		column{2} = {c},
		column{4} = {c},
		column{5} = {c},
		column{6} = {c},
		column{7} = {c},
		column{9} = {c},
		cell{1}{3} = {c},
		cell{1}{8} = {c},
		cell{2}{2} = {r=7}{},
		cell{2}{3} = {c},
		cell{2}{8} = {c},
		cell{3}{3} = {c},
		cell{3}{8} = {c},
		cell{4}{3} = {c},
		cell{4}{8} = {c},
		cell{5}{3} = {c},
		cell{5}{8} = {c},
		cell{6}{3} = {c},
		cell{6}{8} = {c},
		cell{7}{3} = {c},
		cell{7}{8} = {c},
		cell{8}{3} = {c},
		cell{8}{8} = {c},
		cell{9}{1} = {r=2}{},
		cell{9}{2} = {r=2}{},
		cell{9}{3} = {r=2}{c},
		cell{9}{4} = {r=2}{},
		cell{9}{5} = {r=2}{},
		cell{9}{6} = {r=2}{},
		cell{9}{7} = {r=2}{},
		cell{9}{8} = {c},
		cell{9}{9} = {r=2}{},
		cell{11}{2} = {r=7}{},
		cell{11}{3} = {c},
		cell{11}{8} = {c},
		cell{12}{3} = {c},
		cell{12}{8} = {c},
		cell{13}{3} = {c},
		cell{13}{8} = {c},
		cell{14}{1} = {r=2}{},
		cell{14}{3} = {c},
		cell{14}{4} = {r=2}{},
		cell{14}{5} = {r=2}{},
		cell{14}{6} = {r=2}{},
		cell{14}{7} = {r=2}{},
		cell{14}{8} = {r=2}{c},
		cell{14}{9} = {r=2}{},
		cell{16}{1} = {r=2}{},
		cell{16}{3} = {r=2}{c},
		cell{16}{4} = {r=2}{},
		cell{16}{5} = {r=2}{},
		cell{16}{6} = {r=2}{},
		cell{16}{7} = {r=2}{},
		cell{16}{8} = {c},
		cell{16}{9} = {r=2}{},
		cell{18}{1} = {r=2}{},
		cell{18}{2} = {r=2}{},
		cell{18}{3} = {r=2}{c},
		cell{18}{4} = {r=2}{},
		cell{18}{5} = {r=2}{},
		cell{18}{6} = {r=2}{},
		cell{18}{7} = {r=2}{},
		cell{18}{8} = {c},
		cell{18}{9} = {r=2}{},
		hline{1-2,18,20} = {-}{0.08em},
		hline{9,11} = {-}{0.05em},
		}
		work         & Implementation & Abstraction Level      & Location                            & Platform                        & Fault Model             & Application & Programing Language & Data Type           \\
		~\cite{el2019self}       & HW             & RTL                    & Neuron                              & Spiking Neuron Circuit          & Permanent               & SNNs        & N/A                 & N/A                 \\
		~\cite{temam2012defect}       &                & RTL                    & Transistor                          & Spatially Expanded ANN Hardware & Permanent               & Small ANNs  & N/A                 & N/A                 \\
		~\cite{zhang2019fault}       &                & Gate level             & Neuron                              & TPU                             & Permanent               & DNNs        & N/A                 & N/A                 \\
		~\cite{vatajelu2019special}       &                & Behavioral             & Neuron/Synapse                      & Spiking Circuit                 & Permanent               & SNNs        & Python              & N/A                 \\
		~\cite{el2020spiking}       &                & Behavioral             & Neuron                              & Spiking Circuit                 & Permanent               & SNNs        & N/A                 & N/A                 \\
		~\cite{gambardella2019efficient}       &                & Behavioral             & Neuron-Data Path                    & FPGA                            & Permanent               & QNNs        & N/A                 & N/A                 \\
		~\cite{xia2017stuck}       &                & Behavioral             & ReRAM cells                         & ReRAM Accelerator               & Transient               & DNNs        & N/A                 & N/A                 \\
		~\cite{chen2020tensorfi}       & SW/HW         & Behavioral             & Weight/Neuron                       & -                               & Rand/Zero/Bit flip     & General     & Python              & N/A                 \\
		&                &                        &                                     &                                 &                         &             & TensorFlow          &                     \\
		~\cite{hari2017sassifi},~\cite{tsai2021nvbitfi} & SW             & Architectural          & Instruction/ Register Files         & GPU                             & Single/Double bit flip & Any         & Python/Cuda        & N/A                 \\
		~\cite{neggaz2019cnns}       &                & Behavioral             & Weights                             & -                               & Permanent               & CNNs        & Python              & N/A                 \\
		~\cite{li2017understanding}       &                & Behavioral (Algorithm) & Data path/Inside and outside neuron & -                               & Permanent/  Transient    & DNNs        & C++                 & Floating/Fix point \\
		~\cite{reagen2018ares}       &                & Behavioral             & Weights                             & GPU                             & Permanent               & DNNs        & Python              & Fix point           \\
		&                & Application            &                                     &                                 &                         &             &                     &                     \\
		\cite{mahmoud2020pytorchfi}       &                & Behavioral             & Weights/Neuron                      & -                               & Permanent               & DNNs        & Python              & Floating point      \\
		&                &                        &                                     &                                 &                         &             & PyTorch             &                     \\
		ReConFI      & SW/HW         & Behavioral             & Weights/Kernels/bias              & ReRAM Accelerators              & Permanent               & CNNs        & Python              & Floating/Fix point \\
		&                &                        &                                     &                                 &                         &             & PyTorch             &                
	\end{tblr}
\end{adjustbox}
\end{table*}

\section{Conclusion}\label{Conclusion}
Convolutional Neural Networks (CNNs) are widely used in many areas, such as safety-critical systems, which need an efficient and reliable system to run. ReRAM-based PIM accelerators are a promising candidate for such applications due to their non-volatility, high density, low power usage, and, importantly, analog process-in-memory capability. Unfortunately, ReRAM accelerators are prone to transient and permanent faults, which are unavoidable. A hardware and software approach is used in this paper to characterize the impact of permanent faults on ReRAM PIM accelerators that run large-scale CNNs. The fault injection is done in three different phases. In the first phase, the parameter changes through a range. In the second phase, maximum and minimum values are injected into different layers and different parameters of CNNs. In this phase, we observed that maximum and minimum values have different behaviors in different layers. In the third phase, SaH and SaL faults are simultaneously injected into ReRAM and the observations show that with the increasing number of faulty cells, the classification accuracy of CNNs degrades drastically. We illustrated that the output of large-scale CNNs depends on the type and value of faults, topology models, types, location, and the number of each layer, the distance of layers from the output, and types of learning parameters. We also demonstrated that the vulnerability of CNNs accuracy to SaL is significantly higher than SaH. These observations will help to optimize the design of fault-tolerant mechanisms and contribute to develop fault-tolerant ReRAM-based PIM accelerators for CNN algorithms.

%
\bibliographystyle{IEEEtran}
\bibliography{IEEEabrv.bib,mybibfile.bib}

\end{document}